\documentclass[12pt]{JHEP3}
\usepackage{amsmath,amsfonts,amssymb}
\usepackage{epsf}

\newcommand{\be}{\begin{equation}}
\newcommand{\ee}{\end{equation}}
\newcommand{\bea}{\begin{eqnarray}}
\newcommand{\eea}{\end{eqnarray}}
\newcommand{\nn}{\nonumber}
\newcommand{\bra}{{\langle}}
\newcommand{\ket}{{\rangle}}
\newcommand{\tr}{\hbox{ Tr}}

 \newcommand{\myfig}[3]{\begin{figure}[ht]
\begin{center}
\leavevmode \epsfxsize=#2cm \epsfbox{#1}
\end{center}
\caption{#3} \label{fig:#1}
\end{figure}}

\author{ David Berenstein $^{1,\dagger}$, Diego H. Correa$^{2\ddagger}$ and Samuel E. V\'azquez$^{1,\S}$\\
$^1$ Department of Physics, UCSB, Santa Barbara, CA 93106 \\
$^2$ Centro de Estudios Cientif\'{\i}cos, Valdivia, Casilla 1469, Chile
\\
$^\dagger$ \email{dberens@physics.ucsb.edu} $^\ddagger$
\email{dcorrea@cecs.cl}
$^\S$\email{svazquez@physics.ucsb.edu}}

\title{ A study of open strings ending on giant gravitons, spin chains and integrability.}
\abstract{We systematically study the spectrum of open strings
attached to half BPS giant gravitons in the $N=4$ SYM AdS/CFT setup.
We find that some null trajectories along the giant graviton are
actually null geodesics of $AdS_5\times S^5$, so that we can study
the problem in a plane wave limit setup. We also find the
description of these states at weak 't Hooft coupling in the dual CFT. We show how the dual description is given by
an open spin chain with variable number of sites. We analyze this
system in detail and find numerical evidence for integrability. We
also discover an interesting instability of long open strings in
Ramond-Ramond backgrounds that is characterized by having a
continuum spectrum of the string, which is separated from the ground
state by a gap. This instability arises from accelerating the
D-brane on which the strings end via the Ramond-Ramond field. From
the integrable spin chain point of view, this instability prevents
us from formulating the integrable structure in terms of a Bethe
Ansatz construction.}

\preprint{CECS-PHY-06/07}

\begin{document}

\section{Introduction}

The gauge theory/gravity duality is probably one of the deepest ideas in theoretical physics, in that
it gives us in principle the possibility to understand quantum gravity exactly in some setups. The main
tool to address this duality involves the ideas of 't Hooft of the large $N$ expansion of field theories
in terms of a dual string theory \cite{'tH}. A lot of recent progress has happened because there are some
examples where the dual string theory is known. The simplest and most studied example of this duality is
the original formulation of the AdS/CFT correspondence in terms of maximally supersymmetric Yang Mills
theory in four dimensions and type IIB superstring theory compactified on $AdS_5\times S^5$ \cite{malda}.

The setup is such that in the limit where $R$ (the radius of curvature of $AdS_5\times S^5$) is large,
the field theory is strongly coupled in the sense of 't Hooft: $R^4\sim g^2_{YM}N$ is large. Indeed,
it would be very nice to have a good understanding of how this duality works in detail, as it would also
give us hints on how to calculate observables in strongly coupled gauge theories in four dimensions for
more general setups.

There are various avenues to explore this correspondence, based on
different observables that one would want to work with.
Historically, the first set of observables to be addressed were the
dual states to single graviton perturbations of $AdS_5\times S^5$
\cite{WittenAds,GKP} and how to go about testing their correlation
functions. These states are all members of BPS multiplets, so their
energies are protected by supersymmetry.

A few years later, it was discovered that there were interesting geometric limits where the string theory
could be quantized exactly \cite{Blau,Metsaev,Blau2}. Translating the corresponding quantum numbers of
states to the field theory language provided a new large $N$ limit where one also scales the energy and $R$
charge of the observables as one makes the coupling constant large \cite{BMN}. The subsector of states that
one focuses on is closely tied to the supersymmetry of the original system, so the states in questions are
nearly supersymmetric. In these cases, perturbation theory at strong coupling was parametrically suppressed
by the large quantum numbers of the states in question, so it was possible to use perturbation theory to
examine the strong coupling regime of the field theory in a relatively safe environment.

This result produced an interest to make a systematic study  of perturbation theory near the free field limit.
In particular,  it became interesting to compute the full spectrum of anomalous dimensions of all local
operators on the field theory.

In the case where one focuses on operators whose free field
dimension grows at most as $\sqrt N$, the states are roughly
described by a Fock space of closed string states, and the dimension
of the operators in the large $N$ limit is dominated by the planar
diagram expansion. The expansion to one loop order of this problem
revealed a very surprising structure. Minahan and Zarembo
\cite{minahan} discovered that the one loop spectrum of anomalous
dimensions in a subsector of the theory gave rise to an integrable
$SO(6)$ spin chain: a generalization of the Heisenberg  $XXX$ spin
chain. This result was later generalized to include the full set of
local operators of the theory and it was shown that the full planar
one loop spectrum of anomalous dimensions gave rise to an integrable
spin chain model that could be solved by Bethe Ansatz techniques
\cite{BS}. In a parallel development, Bena et. al. \cite{BPR}
discovered that the string sigma model on $AdS_5\times S^5$ was also
integrable giving rise to the idea that the integrability structure
of the string on $AdS_5\times S^5$ could be used as a vehicle to
understand the AdS/CFT correspondence in detail, at least in the
maximally supersymmetric case \cite{DNW}. However, integrability is
not a requisite of the AdS/CFT correspondence, so apart from
circumstantial evidence for integrability at weak and strong
coupling it is not clear that this will be the final answer to the
AdS/CFT puzzle.

Obviously, having an integrable structure is a strong constraint on
the string dynamics and it has been shown that many interesting
general perturbations of the maximally supersymmetric background
destroy the integrability properties (see \cite{BC} for example). In
a similar vein, integrability can also happen for spin chain models
with boundaries and it is interesting to examine if such integrable
structures can also appear in the perturbative Yang Mills theory and
use them as a tool to examine other configurations of matter in the
AdS/CFT correspondence. Indeed, the natural boundaries for open
strings are D-branes, so one can try to see if D-brane solutions
preserve the integrability of the string theory and then try to
solve the string spectrum to understand the D-branes, much like
solving the string spectrum should help to understand gravity.

Here, one can envision two different types of D-branes. The first corresponds to adding defects to the field
theory so that there are ``flavor branes" of various dimensionalities, and the quarks can act as boundaries
for the spin-chain string. These branes are always infinite. This has been explored in \cite{DeWM,CWW,CWW2}, with the
result that the one-loop set of anomalous dimensions are actually integrable.

A second approach is to study finite volume D-branes on the AdS
geometry. These correspond to non-perturbative states of finite
energy in the string description, so one can try to find the dual
description of these states in the field theory. Because of issues
of strong coupling physics, it is better if the corresponding states
are supersymmetric to have some protection for the calculations.
From this point of view, one has to look for supersymmetric D-brane
configurations in the AdS geometry, and to their corresponding dual
description in the field theory. Once these D-branes are found, one
should be able to describe the  open strings attached to them and
study if these strings can be described by a boundary spin chain
model. After this is done, one can ask if the model corresponds to
some form of integrability or not and if this integrability is of
the familiar Bethe-Ansatz form or not. A lot more care is needed in
this case because the branes are finite. This finiteness implies
that the D-brane can back-react to the presence of the string and
the dynamics can be much more complicated than in the case of an
infinite brane.

This is the problem that we will concern ourselves in this paper. The branes under question
are going to be giant gravitons \cite{mcgreeve}. These preserve half of the supersymmetries and
their dual field theory states are known very well \cite{BBNS,CJR,Btoy}. How to add strings to
giants was discussed in the works \cite{BHK,BHLN,david,vijay3}  which also included a description
of how the enhanced gauge symmetry of coinciding D-branes could be understood. One of the main
difficulties in performing the calculations is that the dimension of the operator is of order
$N$ and it is harder to separate the planar and non-planar contributions. The dimension of the
operator implies that there is a combinatorial enhancement of the usual $1/N$ suppressions, so
it is not clear a priori that there is a well defined procedure that works in the general case,
and one has to work example by example to understand the dynamics of the purported D-brane state.

The one loop spectrum of anomalous dimensions for strings attached to a {\em maximal giant graviton}
was described in \cite{sam}, were it was found that the one loop planar anomalous dimensions
correspond to an ordinary spin chain model with integrable Dirichlet-like boundary conditions.
This work was extended to study what spin chain corresponds to a more general giant graviton in
\cite{bcv}, where we found that the spin chain in question has a variable number of sites and
therefore it is not an ordinary spin chain model anymore. After a bosonization transformation,
we found that the spin chain model could be also understood  in terms of a system of a Cuntz
oscillator chain model (a boson chain, where each spin corresponds to a single boson Fock
space) with non-diagonal boundary conditions. The non-diagonal boundary conditions imply that
the total boson number is not conserved. Our setup is also the limit $q\to 0$ of a $q$-boson chain
model with non-diagonal boundary conditions, where the $q$ boson is defined by the algebra
$a a^\dagger-qa^\dagger a=1$, with $q$ a real number. More recently, it has been claimed by
\cite{Agarwal} that in the {\em maximal giant graviton} case, there seems to be a problem with
integrability beyond the two loop order, in contrast to the closed spin chain model \cite{AFS,BS2}.
Agarwal argued that the Bethe Ansatz breaks down by direct calculation at two loops and therefore the model is not integrable. Another claim that studies the consistency of a Bethe Ansatz for the maximal giant graviton by Okamura and Yoshida suggest that the BMN limit breaks down instead \cite{OY}.
We have a different interpretation of these facts: if the system is integrable, it will not realize integrability by a Bethe Ansatz. To do this, we study more general giant gravitons.

In this paper we extend our analysis of \cite{bcv} to try to understand the full spectrum of the
variable length spin chain model. The hamiltonian of the spin chain model is given by 
\begin{eqnarray}
 H &=& 2
\lambda \sum_{l=  1}^L \hat{a}_l^\dagger \hat{a}_l - \lambda
\sum_{l=1}^{L-1} (\hat{a}_l^\dagger\hat{a}_{l+1} +
\hat{a}_l \hat{a}_{l+1}^\dagger)
\nonumber \\ && +  2 \lambda \, \alpha^2
+ \lambda \,\alpha \,(\hat{a}_1^\dagger + \hat{a}_1) +
\lambda\, \alpha\,(\hat{a}_L^\dagger + \hat{a}_L)\;,
\end{eqnarray}
and the operators  ${\hat a}_i$ obey
the Cuntz algebra for a single species,
\begin{eqnarray}
\label{ops} {\hat a}_i {\hat a}_i^\dagger = I\;, \quad {\hat
a}_i^\dagger {\hat a}_i = I - |0\ket\bra 0|\;,
\end{eqnarray}
where operators corresponding to different sites commute. \footnote{Since spin chain models also have applications to other areas of physics, the reader who is only interested in the analysis of the spin chain model can jump to section \ref{dual} and should also read sections \ref{sec:unstable},\ref{sec:integrability} }

 In our exploration of the model we will find very interesting phenomena that at first
sight seem to contradict a very naive intuition about the system. The naive intuition is that the open
string ending on the D-brane has a discrete spectrum, because one has a finite energy configuration of
matter in the $AdS_5$ geometry. However, there is a technical issue with the fact that the energy is
going to infinity as we take $N$ large, because the mass of the D-brane scales with $N$. The naive answer,
in spite of being technically correct, might also include some non-perturbative information on the
finiteness of $N$, that is not necessarily reflected in the strict large $N$ limit. Indeed, we find that
the spectrum of the spin chain model is continuous. Upon thinking further about it, it becomes obvious
that the spin chain answer correlates very well with the AdS geometric intuition and we will explain
how this puzzle is solved in detail.

The continuous string spectrum also has some other consequences for the integrability program. It seems to
indicate that if integrability is present, it will not be of the form of a Bethe ansatz solution. From this
point of view, if integrability is present, the conjectures in \cite{BS2} need to be extended appropriately, or it is even possible that they might need to be revised completely.
For finite size systems, a Bethe ansatz solution would predict that the spectrum of the system is related to
solutions (roots) of polynomial equations of high degree, giving rise to a discrete spectrum of
configurations. Does this fact destroy the integrability of the open string? We don't think so. Although it
is hard to come by with integrable systems that can not be solved by Bethe Ansatz, there are some well known 
examples of integrable systems that manage to be integrable without having a Bethe ansatz solution. Indeed,
the $c=1$ matrix model is such an example. This model is dual to a two dimensional string in a linear
dilaton background. It is very well known that the number of particle excitations is not conserved when
scattering of the Liouville wall, so this non-conservation of the number of particle excitations is contrary
to the typical Bethe ansatz solution. There are other hints that a Bethe ansatz is not the complete story for
describing the individual string dynamics at strong 't Hooft coupling \cite{BCV}. We believe our results in
this paper show that the Bethe ansatz breaks down perturbatively a lot faster for the open string than for the
closed string.

Our paper is organized as follows. We begin in section \ref{OSGG} by reviewing the physics of giant
gravitons. We show how the local geometry near an open string spinning around the moving giant
graviton arises from a Penrose limit of $AdS_5 \times S^5$. We quantize some of the bosonic modes
of the open string in that background, which is valid for short strings that sense the local geometry.
We then consider longer strings that carry two angular momenta, one of them in the same direction as
the giant graviton. Then we expand the Polyakov action in the limit of large transverse angular
momentum just like with closed strings \cite{Kruczenski, kru}.

In section \ref{dual} we move to the dual description in SYM theory. We review the derivation of
the matrix of anomalous dimensions interpreted as the Hamiltonian of a spin chain with variable number
of sites. We then map this problem to a lattice of bosons with sources and sinks at the boundaries.
Then, we show numerical evidence for the agreement of the bosonic Hamiltonian spectrum with the plane wave
spectrum of the dual open string. We also use coherent states to obtain a sigma model action in the
limit of large R-charge. It agrees with the Polyakov action of the open string in the large
momentum limit, in a particular gauge. Moreover we show how the SYM theory gives the correct boundary
conditions for the open strings.

In section \ref{sec:unstable} we show that the spectrum of the anomalous dimension matrix contains
continuum energy bands. We then argue that these can represent an instability of the dual D-brane
where the open string can absorb a significant fraction of the angular momentum of the giant, making
it collapse. In fact, we show that the lowest energy band is accessible to semiclassical string excitations.

In section \ref{sec:integrability} we show some numerical evidence that lead us to conjecture the integrability
of the Hamiltonian. Finally, in section \ref{sec:discussion}, we discuss our results and a list of open
problems that hopefully will be of interest for the reader.  Some of the details of the calculations for the
different sections are shown in the appendices.

\section{Open Strings on giant gravitons in $AdS_5 \times S^5$}
\label{OSGG}

In this section we study giant gravitons in $S^5$ and their open
string excitations  from the point of view of string
theory/supergravity.  In the next section we will see how this
stringy description arises from the dual SYM theory.

We begin by briefly reviewing some of the basic geometrical
properties of the spherical D3-branes existing in the background of
$AdS_5\times S^5$ that wrap an $S^3$ inside the $S^5$
\cite{mcgreeve}. These spherical branes are commonly called giant
gravitons.

They are 1/2 BPS and their classical stability is due to the
presence of the five-form flux which exactly cancels their tension.
As we will see, the movement of the non-maximal giant gravitons
gives rise to non-trivial boundary conditions to open string
excitations. We are interested in studying fast moving (nearly
supersymmetric) strings on the D-brane (close to the speed of light
and near the point particle limit), as for these strings the
dynamics simplifies. These should be closely related to null
trajectories in $AdS_5\times S^5$ that are contained in the D-brane
worldvolume trajectory. Moreover,  as we now show, for the special
supersymmetric case the geometry near this null trajectory  is the
usual pp-wave of type IIB supergravity, so we will focus on this
case.

\subsection{Spherical D3-branes in $AdS_5\times S^5$}
Spherical stable branes are known to exist in the maximally supersymmetric
background of type IIB supergravity $AdS_5\times
S^5$ \cite{mcgreeve}. Indeed, these branes are actually supersymmetric. 
They can wrap an $S^3$ either inside the $S^5$
or inside of $AdS_5$ \cite{grisaru, Hashimoto}.
In this section and in the rest of the paper we will focus on the former ones.

To begin with, let us write the metric of the  $AdS_5\times S^5$
background as
\be \label{ori} ds^2 = R^2(-\cosh ^2 \rho dt^2 +d\rho^2 +\sinh ^2
\rho d{\Omega'}_3^2  +
             d\theta^2 + \cos ^2 \theta d\psi^2 + \sin^2\theta d{\Omega}^{2}_3)\, ,
\ee
 and the 3-sphere metrics as
\bea &&d{\Omega}^{2}_3 \ = d\varphi^2 + \cos^2\varphi d\eta^2
+\sin^2\varphi d\xi^2\, ,\nn\\
&&d{\Omega'}^{2}_3 = d{\varphi'}^2 + \cos^2{\varphi'} d{\eta' }^2
+\sin^2{\varphi'} d{\xi'}^2\, . \eea
In these coordinates, the Ramond-Ramond 4-form potential takes the
form \be C^{(4)} = 4 \pi N {\alpha'}^2\left(\sinh^4\rho\
dt\wedge\Omega'_3 - \sin^4\theta\ d\psi\wedge\Omega_3\right)\, , \ee
and we have the relation $R^4 = 4\pi g_s N \alpha'^2$.

McGreevy, Sussking and Toumbas found spherical D3-branes carrying
angular momentum in the $S^5$ \cite{mcgreeve}. They are
supersymmetric solutions of the brane action, that
expand in the 3-sphere $\Omega_3$. Choosing
the static gauge, the parametric coordinates of the brane
$(\sigma_0,\sigma_1,\sigma_2,\sigma_3)$ can be identified with
space-time coordinates
\be t=\sigma_0\, , ~ ~ ~\varphi=\sigma_1\, ,
 ~ ~ ~  \eta=\sigma_2\, ,  ~ ~ ~ \xi = \sigma_3\, .
\ee
The brane carries angular momentum along the
$\psi$ direction and it is located at  $\rho=0$ and at a constant
$\theta=\theta_0$.
\be \psi=\psi(\sigma_0)\, ,  ~ ~ ~ ~\rho=0\, , ~ ~ ~ ~ ~
\theta=\theta_0 \, , \ee
The equations of motion are solved with $\dot\psi$ constant. More
precisely, and independently of $\theta_0$, by $\dot\psi=1$. Thus,
the center of mass of the giant graviton is moving along an
equatorial null trajectory. However, each element of the giant is
moving in a time-like orbit of radius $r_{el}=R\cos\theta_0$. The
radius of the giant also depends on $\theta_0$,
 $r_{gg} = R\sin\theta_0$.

For $\dot\psi=1$, the momentum conjugate to $\psi$ becomes
\be
 p = N \sin^2\theta_0\,.
 \label{p}
\ee
We can use this relation for highlighting a relevant feature of
giant gravitons: their radii  grows as they increase their angular
momenta \be r_{gg}= R \sqrt {\frac pN}\, . \label{rgg} \ee Since we
are considering giant gravitons that expand in $S^5$, the radius is
bounded by $r_{gg}\leq R$ and hence the angular momentum is also
bounded by the number of units of five-form flux on the sphere \be p
\leq N\, . \ee When the equality is satisfied  the brane solution is
known as {\em maximal giant graviton}. Notice that in this case the
radius of the orbit of an element of the giant $r_{el}=\sqrt{1-p/N}$
shrinks to zero and all the angular momentum comes from the term
with the Ramond-Ramond form in the brane action. Being the maximal
giant graviton static is the reason why this special case turns out
to be simpler. However, considering {\em non-maximal giant
gravitons} gives rise to interesting and novel phenomena.

\subsection{A Penrose limit for an open string on a non-maximal giant}

We obtain in this section, the effective geometry seen, in the large
$N$ limit, by an open string with angular momentum $J$ growing as
$\sqrt{N}$, attached to a giant graviton whose angular momentum $p$
is growing proportionally to $N$. Then, the angle $\theta_0$ is kept
constant and the radius of the giant diverges. The open string will
be effectively attached to a flat D3-brane in a pp-wave background.
For the maximal giant graviton this was already analyzed in
\cite{BHLN}. We now focus on smaller or non-maximal giant gravitons.

As we already said, each element of the giant graviton is moving in
a time-like trajectory. In order to travel in null trajectory, an
observer on the giant should be spinning fast along it. Then, we
will study open strings with two components of angular momenta. In
particular, we now consider a trajectory along $\psi$ and $\eta$,
keeping $\varphi =0$. A null trajectory should satisfy
\be R^2(- {\dot t}^2 +\cos^2\theta_0  {\dot \psi}^2+ \sin^2\theta_0
{\dot \eta}^2)=0\, . \label{null} \ee
Since ${\dot t}=1$ and ${\dot \psi}=1$, one necessarily has
$\dot\eta= \pm1$. Although it is not obvious at first sight, this
null trajectory is a null geodesic of $AdS_5\times S^5$ and then a
customary Penrose limit can be taken. At this point it is worth
highlighting the ratio of the angular momentum in both angular
directions
\bea \label{ratio} \frac{J_\psi}{J_\eta}=\cot^2 \theta_0 =
\frac{N}{p}\left(1-\frac{p}{N}\right). \eea
We will later see the natural emergence of this quantity in the
gauge theory side.

By following the standard Penrose limit procedure (see appendix
\ref{apl} for details) we get a plane-wave geometry with the metric
\be ds^2 = -4dudv + 4 y dudx - \sum^6_{a=1} z_a^2 du^2 +dx^2+dy^2+
\sum^6_{a=1} dz_a^2  \, . \label{ppw} \ee
The RR 5-form field strength becomes
\be F_{(5)} = 2 du\wedge(dz_1\wedge dz_2 \wedge dz_3 \wedge dz_4 +
dz_5\wedge dz_6 \wedge dz_7 \wedge dz_8 )\, . \ee
This is nothing but the usual maximally supersymmetric pp-wave of
type IIB supergravity \cite{Blau} displayed in unusual coordinates.
This can be explicitly seen with an appropriate coordinate
transformation \cite{Miche, Bertolini}.

\subsection{Open Strings on the pp-wave Geometry (Short Strings)}

Let us now consider the open string theory in the pp-wave geometry
(\ref{ppw}) corresponding  to an open string spinning along a
non-maximal giant.  This is a good description for short strings
which sense the local geometry near the giant graviton.  We will
focus on the bosonic sector of the superstring and, in particular,
on those modes that later will be compared to the gauge theory
predictions. More precisely, we will focus on open strings
carrying two angular momenta on the sphere with one of them in the
direction of movement of the giant graviton.

The bosonic part of the superstring action is
\bea
S\!&=&\!-\frac{1}{4\pi\alpha'}\int\! d\tau\!\int_0^\pi\! d\sigma\left(
-4\partial^\alpha u\partial_\alpha v + 4 y \partial^\alpha u\partial_\alpha x
- z_i z_i \partial^\alpha u\partial_\alpha u \right. \nonumber
\\ &&\;\;\;\;\;\;\;\;\;\;\;\;\;\;\;\;\;\;\;\;\;\;\;\;\;\;\;\;\;\;\;\left.
+ \partial^\alpha x\partial_\alpha x +\partial^\alpha y\partial_\alpha y
+\partial^\alpha z_i\partial_\alpha z_i\right)\, .
\eea
Fixing the light-cone gauge with the usual procedure
\be
u=2 \alpha' p^{u}\tau\, ,
\label{lcu}
\ee
we obtain the light-cone action
\be\label{lc}
S_{lc}=-\frac{1}{4\pi\alpha'}\int\! d\tau\!\int_0^\pi\! d\sigma\left(
\partial^\alpha x\partial_\alpha x +\partial^\alpha y\partial_\alpha y
-4m y  \dot x
+\partial^\alpha z_i\partial_\alpha z_i  + m^2 z_i z_i  \right)\, ,
\ee
where dots and primes refer to derivatives with respect to $\tau$
and $\sigma$ respectively. We have also defined the mass $m= 2
\alpha' p^{u}$. As in the case of closed strings \cite{FT}, this
action can also be derived as the quadratic order expansion of the
string action around the classical solution $\theta=\theta_0$ and
$\psi=t$.

The equations of motion from (\ref{lc}) are
\bea
(\partial^\alpha\partial_\alpha - m^2) z_i = 0\, , && \\
 \partial^\alpha\partial_\alpha x + 2 m \dot y =0\, ,&&\\
 \partial^\alpha\partial_\alpha y - 2 m \dot x =0\, .&&
\eea
For $z_i$ we obtain the usual massive equations. For $x$ and $y$
it is convenient to define a complex field $w=x +i y$ which
satisfies
\be
\partial^\alpha\partial_\alpha w - 2i  m \dot w =0\, ,
\label{w}
\ee
A solution of (\ref{w}) is $w=e^{-im\tau} W$ whenever $W$ is a solution
of the massive equation
\be \label{ww} \left(\partial^\alpha\partial_\alpha   - m^2\right)
W =0\, . \ee
We now have  to specify the boundary conditions for these fields.
To this end, we have to keep track  of the  D-brane position in
the Penrose limit. This can be done using the coordinate
transformation of the Penrose limit (see appendix \ref{abpl}). The
original boundary conditions  are translated into Neumann boundary
conditions for $u,v,z_1,z_2$ and Dirichlet boundary conditions for
$x,y,z_3,z_4,z_5,z_6$.

The mode expansion and the canonical quantization of fields $z_1,\ldots,z_6$ (those
satisfying massive equations of motion), goes exactly as in \cite{DP}.  We  concentrate on
excitations of fields $x$ and $y$ which later will be related to the dual description.

Solutions of (\ref{ww}) satisfying Dirichlet boundary conditions can be expanded as
\be
W = \sum_{n> 0} \sin(n\sigma)\sqrt{\frac{4\alpha'}{\omega_n}}
\left(\beta_n e^{-i \omega_n\tau}+\tilde\beta_n^{*} e^{i \omega_n\tau}\right)\, ,
\ee
where $\omega_n = +\sqrt{n^2 +m^2}$. The factor $\sqrt{{4\alpha'}/{\omega_n}}$
in the coefficients is included for later convenience. The expansions for the original fields
$x$ and $y$ are
\bea x&=&\frac{1}{2} \sum_{n> 0}
\sin(n\sigma)\sqrt{\frac{4\alpha'}{\omega_n}} \left(\tilde\beta_n
e^{-i \omega^{-}_n\tau}+\beta_n e^{-i \omega^+_n\tau}
+\tilde\beta_n^{*} e^{i \omega^{-}_n\tau}+\beta_n^{*} e^{i
\omega^+_n\tau}\right)\;,
\\
y&=&\frac{i}{2} \sum_{n> 0} \sin(n\sigma)\sqrt{\frac{4\alpha'}{\omega_n}}
\left(\tilde\beta_n e^{-i \omega^{-}_n\tau}-\beta_n e^{-i \omega^+_n\tau}
-\tilde\beta_n^{*} e^{i \omega^{-}_n\tau}+\beta_n^{*} e^{i \omega^+_n\tau}\right)\;,
\eea
where now, $\omega_n^{\pm} = \omega_n \pm m$. The string spectrum can be
obtained by canonical quantization. With our normalization, the coefficients
upgraded to operators satisfy two set of mutually commuting oscillator-like algebra,
\bea
&&[\beta_n,\beta_m^\dagger] =\delta_{nm}\;,\;\;\;\;\;\;\;\;\;\;\;\;\;\
[\beta_n,\beta_m] =   [\beta_n^\dagger,\beta_m^\dagger] =0\; ,\\
 &&[\tilde\beta_n,\tilde\beta_m^\dagger] =\delta_{nm} \;,\;\;\;\;\;\;\;\;\;\;\;\;\;\
[\tilde\beta_n,\tilde\beta_m]= [\tilde\beta_n^\dagger,\tilde\beta_m^\dagger] =0\; .
\eea
The spectrum of the sector we are considering is obtained by acting with the creation
operators $\beta_m^\dagger$ and $\tilde\beta_m^\dagger$ on a vacuum state
satisfying $\beta_m|0\rangle=\tilde\beta_m|0\rangle=0$. The light-cone Hamiltonian
can be expressed in terms of the number operators corresponding to the oscillator operators.
Scaling $\tau$ and $\sigma$ by $2\alpha'p^u$
\bea
H_{lc}^{xy}\!&=&\!\frac{1}{8{\alpha'}^2 p^u}\int_0^{2\pi\alpha'p^u}\!\!\!\!d\sigma
\left(\dot x^2 +\dot y^2 + {x'}^2+{y'}^2\right)
\nn\\
&=& \frac{1}{2{\alpha'} p^u}\sum_{n>0}
\left(\omega_n^-\tilde\beta_n^\dagger\tilde\beta_n+\omega_n^+\beta_n^\dagger\beta_n\right)\;.
\eea
We see that excitations created by $\beta_n^\dagger$ have  more
energy than those created by $\tilde\beta_n^\dagger$. Expanding
the square roots in $\omega_n$
\be H^{xy}_{lc}\approx\sum_{n>0} \left[ \tilde N_n
\frac{n^2}{8{\alpha'}^2 {p^u}^2} +
N_n\left(2+\frac{n^2}{8{\alpha'}^2 {p^u}^2}\right)\right]\, , \ee
where $N_n = \beta_n^\dagger\beta_n$ and $\tilde N_n = \tilde\beta_n^\dagger\tilde\beta_n$.

~

The Hamiltonian and angular momentum generators are, in the light-cone gauge,
\bea
H_{lc} \!&=&\! -p_u = i\frac{\partial}{\partial u}\, ,\\
p^u\!&=&\! -\frac{1}{2}p_v=\frac{i}{2}\frac{\partial}{\partial v}\,.
\eea
Using the change of coordinates (\ref{cha}) we can express them in terms
of the original generators
\bea
H_{lc} &=& i\left(\frac{\partial}{\partial t}+ \frac{\partial}{\partial \psi}
+\frac{\partial}{\partial \eta}\right)= \Delta - J_\psi -J_\eta\, ,\\
p^u &=& \frac{i}{2 R^2}\left(\frac{\partial}{\partial t}- \frac{\partial}{\partial \psi}
-\frac{\partial}{\partial \eta}\right)=\frac{\Delta + J_\psi +J_\eta}{2R^2}\,.
\eea

Looking forward to a gauge theory interpretation we can label the
angular momenta as (c.f. (\ref{ratio})),
\bea
J_\eta &=& L\, ,\\
J_\psi &=& \cot^2\theta_0 L =\frac{\alpha^2}{1-\alpha^2}L\,.
\eea
where we define $\alpha \equiv \sqrt{1 - p/N}$.  Then, the sum of
angular momenta is
\be
J_\eta+J_\psi = \frac{L}{1-\alpha^2}\;.
\ee
For a finite light-cone energy it is required that
$\Delta\simeq L/(1-\alpha^2)$ and then $p^u\simeq L/R^2(1-\alpha^2)$.
Finally, using $R^4 = 4\pi g_s N \alpha'^2$ and $\lambda=g_s N/2\pi$, the
energy of each excitation is
\be \label{ppspec} \tilde E_n \approx \frac{\lambda \pi^2
(1-\alpha^2)^2 n^2 }{L^2}\, , \;\;\;\;\;\;\;\; E_n \approx 2+
\frac{\lambda \pi^2 (1-\alpha^2)^2 n^2 }{L^2}\;, \ee for  $x$ and
$y$ respectively. Thus, as expected,  we see that string theory
predicts a BMN limit for the anomalous dimension of the dual
operators describing these open string excitations.  We will come
back to the dual interpretation of these energies in section 3.

\subsection{Semiclassical Open Strings (Long Strings)}

If we want to consider more general open strings ending on the
giant graviton, we need to include the full $AdS_5 \times S^5$
background in the Polyakov action. However, it is well known by
now that we should expect a classical description in the large
angular momentum limit\footnote{The literature on this subject is
very extensive, but for a nice review see \cite{Tseytlinrev}.}.
 Again, as we did
for short strings in the pp-wave geometry, we focus on the
excitation of two coordinates of the string. More precisely,  the
two coordinates of the sphere $S^5$ subject to Dirichlet boundary
conditions. The reason is that the description of these
excitations in the dual theory is more easily isolated from the
rest. Moreover we will fix a particular gauge of reparametrization
invariance different from the standard conformal gauge, following
closely \cite{kru}. This election will be the reflection of the a
particular labeling of the operators in the dual gauge theory.

These results were already presented in a previous letter
\cite{bcv}. We reproduce them here providing more details of their
derivation. The starting point is the Polyakov action in phase
space. We can write the conjugate momenta as,
\be
p_\mu= -G_{\mu\nu}(A\partial_0 x^\mu + B \partial_1 x^\mu)\, ,
\ee
where $A=\sqrt{-g} g^{00}$, $B=\sqrt{-g} g^{01}$ and $g^{ab}$ is
the worldsheet metric. The Polyakov action then takes the form
\begin{eqnarray}
\label{paction} S_p = \sqrt{\lambda_{YM}} \int d\tau \int_0^\pi
\frac{d\sigma}{2 \pi} {\cal L}\;,
\end{eqnarray}
where,
\begin{eqnarray}
{\cal L} &=& p_\mu \partial_0 x^\mu + \frac{1}{2} A^{-1} \left[
G^{\mu \nu} p_\mu p_\nu + G_{\mu \nu} \partial_1 x^\mu
\partial_1 x^\nu \right] + B A^{-1} p_\mu \partial_1 x^\mu\;.
\end{eqnarray}
Here we have factorized the radius of $AdS_5$ and $S^5$ so that by
the AdS/CFT correspondence  $\lambda_{YM} = g_{YM}^2 N = 8 \pi^2
\lambda =  R^4/\alpha'^2$. Moreover,  $A$, $B$ play the role of
Lagrange multipliers implementing the constraints $T_{a b} = 0$.

For an open string traveling with the giant graviton and with excitations
on the sphere only, the effective geometry is $\mathbb{R} \times S^5$.
We can write the corresponding metric as,
\begin{eqnarray}
ds^2 = -dt^2  + |dX|^2 + |dY|^2 + |dZ|^2\;,
\end{eqnarray}
where $|X|^2 + |Y|^2 + |Z|^2 = 1$. The giant graviton will be
orbiting in the $Z$ direction with $Z = \sqrt{1 - p/N} e^{i t}$
and will wrap the remaining $S^3$. We will put our string at $X =
0$. Then we define the coordinates,
\begin{eqnarray}
Z &=& r e^{i (t  - \phi)}\;,\\
Y &=& \pm \sqrt{1 - r^2} e^{i \varphi}\;,
\end{eqnarray}
for which the giant graviton is static at $r = \sqrt{1 - p/N}$
and, say, $\phi = 0$ (we can always shift $\phi$ by a constant).
The metric becomes,
\begin{eqnarray}
\label{metric}
 ds^2 &=&  -(1 - r^2) dt^2 + 2 r^2 dt d\phi
+ \frac{1}{1 - r^2} dr^2  + r^2 d\phi^2 + (1 - r^2) d\varphi^2 \;.
\end{eqnarray}
The connection between these coordinates and the plane wave
coordinates is shown in appendix \ref{apl}.

The momentum in  $\varphi$ is conserved and is given by,
\begin{eqnarray}
 L  = \sqrt{\lambda_{YM}} \int_0^\pi \frac{d\sigma}{2 \pi}
p_\varphi \equiv \sqrt{\lambda_{YM}} {\cal J} \;.
\end{eqnarray}
  We choose a gauge that
distributes the
 angular momentum  $p_\varphi$
homogeneously along the string. Furthermore, we choose $\tau$ to
coincide with the global time in the metric. Thus, our gauge is
\begin{eqnarray}
\label{gauge}
t = \tau \;, \;\;\;\;\;\; p_\varphi = 2 {\cal J} = {\rm const} .
\end{eqnarray}
These type of gauges were introduced in \cite{Aru}

We then implement the constraints that follow from varying $A$ and
$B$ in (\ref{paction}) directly in the action. This allows us to
write the Lagrangian in terms of the momenta $p_r$ and $p_\phi$
and the fields $r$ and $\phi$ and their derivatives. These  are
the two fields subject to Dirichlet boundary conditions. We get
(up to a total derivative in $\tau$),
\begin{eqnarray}
\label{Lagmom} {\cal L} = (1 - \dot{\phi})\, p_\phi + \dot{r}\,
p_r  - \sqrt{a \, p_\phi^2 + b\, p_r^2 + 2 c\, p_r\, p_\phi +
d}\;,
\end{eqnarray}
where,
\begin{eqnarray}
a &=& 1 + (1 - r^2) \left( \frac{{\phi'}^2}{4 {\cal J}^2}  +
\frac{1}{r^2} \right)\;,\\
b  &=& (1 - r^2) \left( \frac{{r'}^2}{4 {\cal J}^2}  +
1 \right)\;,\\
c &=& - \frac{{r'} {\phi'} (1 - r^2)}{4 {\cal J}^2}\;, \\
d &=& \frac{4 {\cal J}^2}{1 - r^2} + r^2 {\phi'}^2 +
\frac{{r'}^2}{1 - r^2}\;.
\end{eqnarray}
As usual, dots and primes denote derivatives with respect to
$\tau$ and $\sigma$ respectively. The remaining gauge freedom can
be fixed by demanding that the equations of motion for $p_r$,
$p_\phi$, $r$ and $\phi$ that follow from (\ref{Lagmom}) agree
with the ones derived from the original action (\ref{paction})
\cite{kru}. This will fix the value of $B$ in terms of $p_r$,
$p_\phi$, $r$ and $\phi$.

Since the momenta $p_r$ and $p_\phi$  enter the Lagrangian
(\ref{Lagmom}) algebraically, we can solve for them using their
equations of motion and write the action in terms of the fields
$r$ and $\phi$ and their derivatives.  We get,
\begin{eqnarray}
{\cal L} = -\left( \frac{d}{1 - a\, {\tilde p_\phi}^2  - b \,
{\tilde p_r}^2 + 2 c\, {\tilde p_\phi} {\tilde p_r} }\right)^{1/2}
\left[1  + (1 -\dot{\phi}) {\tilde p_\phi}- \dot{r}\, {\tilde
p_r}\right]\;,
\end{eqnarray}
where,
\begin{eqnarray}
{\tilde p_\phi} &=& \frac{\dot{r} c - (1-\dot{\phi}) b }{a b - c^2}\;, \\
{\tilde p_r} &=& \frac{\dot{r} a - (1- \dot{\phi}) c }{a b -
c^2}\;.
\end{eqnarray}

Finally, we take the limit ${\cal J} \rightarrow \infty$ and assume that the
time derivatives are of the order $\partial_0 x^\mu \sim 1/{\cal
J}^2$. This last condition can be made more precise by solving for
the time derivatives in terms of the spatial derivatives as in
\cite{kru}. However, in our case this is not necessary because we
just want the lowest order in $1/{\cal J}$. Thus, rescaling
$\sigma \rightarrow \pi \sigma$, we get
to lowest order,
\bea \label{stringaction}
S \approx -L \int dt \int_0^1 d\sigma
\left[\frac{r^2 \dot{\phi}}{1 - r^2} + \frac{\lambda}{L^2} (r'^2 +
r^2 \phi'^2) + {\cal O}\left(\frac{\lambda^2}{L^4} \right)\right]\;. \eea
We  note that
$L$ serves as an inverse ``Planck constant" and so $L \rightarrow
\infty$ corresponds to a classical limit as promised. This coordinates
are subject to Dirichlet boundary condition, which are expressed as,
\begin{eqnarray}
\label{bc}
r|_{\sigma = 0, 1} &=&  \sqrt{1 - \frac{p}{N}}\;, \\
\phi|_{\sigma = 0, 1} &=& {\rm const.}
\end{eqnarray}

\section{Open Strings on Giant Gravitons from ${\cal N} = 4$ SYM}
\label{dual}

In this section we study the dual description of open
strings on giant gravitons given by ${\cal N} = 4$ SYM. It
is very interesting to see how the complete geometrical picture
discussed in the last section is recovered from the gauge theory.
In particular, we show the emergence of the BMN
spectrum, the Polyakov action, the Dirichlet boundary conditions
and the reparametrization invariance of the string world sheet. All
of these geometrical ingredients are encoded in the matrix of
anomalous dimension of the operators dual to  giant gravitons with
open string excitations.

There is considerable evidence that the dual
operators to these D3-branes are of determinant-like form
\cite{BBNS, CJR, david} (see appendix \ref{con}
for our conventions)
\be \label{ope} { \cal O}^p = \epsilon_{i_1 \cdots i_p}^{j_1
\cdots j_p} Z_{j_1}^{i_1} \cdots Z_{j_{p}}^{i_{p}}\;, \ee
where $Z$ is one of the three complex scalars of ${\cal N} = 4$
SYM. These are chiral operators and, by supersymmetry, their
dimensions are determined in terms of their R-charge:  $\Delta =
J$ (BPS condition). For ${ \cal O}^p$ we see that $\Delta = p \leq
N$, and hence this operator obeys the ``momentum" bound of the
giant gravitons. This makes sense from the AdS/CFT correspondence
since we should identify the $U(1)$ charge of (\ref{ope}) as the
angular momentum of the D-brane and $\Delta$ as its energy.

The operator dual  to a single string  attached to an $S^5$ giant
graviton of momentum $p$ is obtained by appending a word to the
determinant-like operator \cite{BHLN, vijay3}, which represents
the open string \footnote{More precisely, as we will see later,
the dual operator to the {\it classical} non-maximal giant
graviton is actually a coherent state of operators like (\ref{op1})
with different values of $p$.},
\begin{eqnarray}
\label{op1} { \cal O}_W^p = \epsilon_{i_1 \cdots i_p}^{j_1 \cdots
j_p} Z_{j_1}^{i_1} \cdots Z_{j_{p-1}}^{i_{p-1}}  W_{j_p}^{i_p}\;,
\end{eqnarray}
where $W$ is a ``word" built out of the scalar fields $X, \bar{X},
Y, \bar{Y}, Z, \bar{Z}$, as long as we only consider excitations on
the $S^5$ directions. Each scalar field carries a unit of $U(1)$
charge corresponding to a unit of angular momentum in one of the
three orthogonal planes that cut $S^5$ in the dual string picture.

We also have the condition that the $Z$ field is not allowed to be
at the borders of $W$ \cite{vijay3}. This constraint comes from
the fact that when a $Z$ is at the border of $W$, the operator
factorizes as a bigger giant plus a giant of the same size with a
closed string (see appendix \ref{con}). In fact, in general, the
operator (\ref{op1}) can be expanded in terms of traces
\cite{david}. However, for $p$ sufficiently large the mixing of
(\ref{op1}) with closed strings (traces) is suppressed in the large
$N$ limit. For this reason we will only consider operators with
$\sqrt{N} \lesssim p \leq N$.

\subsection{Open Strings as Variable Length Spin Chains}
We are interested in computing the mixing matrix of anomalous
dimension for this type of operators at the one-loop approximation
and large $N$
 limit. Thus, we will only consider planar diagrams. As
usual, we begin by defining the correlation function
\begin{eqnarray}
M_{A B} = \left \langle \tilde{\cal O}^*_A(x) \tilde{\cal O}_B(0)
\right \rangle_{\text{free + interacting}} \;,
\end{eqnarray}
where the operators have been normalized according to
\begin{eqnarray}
\tilde{\cal O}_A(x) = \frac{{\cal O}_A(x)}{\left \langle {{ \cal
 O}}_A^*(x) {\cal O}_A(0) \right \rangle_{\text{free}}^{1/2}}\;,
\end{eqnarray}
and where $A$, $B$ are collective indices labeling different giant
graviton configurations with a single string. At one-loop and in
the large $N$ limit, $M_{A B}$ will have the following
general form,
\begin{eqnarray}
\label{Mexp}
 M_{A B} = \frac{1}{|x|^{2\Delta_0}}(\delta_{A B} -
2 \, \Gamma_{A B} \log(|x|\Lambda) + \ldots) \;,
\end{eqnarray}
where $\Delta_0$ is the classical dimension, $\Gamma_{A B}$ is the matrix of
anomalous dimension and $\Lambda$ is an ultraviolet cutoff. We then identify the
anomalous dimension matrix with the Hamiltonian of the corresponding string
quantum states,
\begin{eqnarray}
\label{gamma}
 \Gamma_{A B} \cong \langle \psi_A
|H|\psi_B \rangle\;,
\end{eqnarray}
where $\tilde{\cal O}_A \cong |\psi_A \rangle$. This is just the
operator state correspondence which is available for any CFT: the
Hamiltonian corresponds to taking radial time and compactifying
the CFT on a round sphere.

In \cite{sam} the case of a maximal giant graviton was considered
and it was shown that the resulting planar anomalous dimension matrix
corresponds to an integrable open spin chain with $SO(6)$ symmetry.
For the case of non-maximal giants, most of the field contractions
are the same as with the maximal giants, the difference being the
particular combinatorics of the $\epsilon$ symbol for $p \neq N$.
However as we remarked in \cite{bcv}, there is a very important new
interaction in the case of the non-maximal giant graviton: $Z$
fields can be exchanged between the word $W$ and the rest of the
operator. This accounts for the exchange of angular momentum between
the giant and the string since the open string is ``dragged" by the
non-trivial movement of the non-maximal giant graviton.

The details of the combinatorics involved for this and the rest of
the interactions between the fields in the operator are explained in
the appendix \ref{con}. Here we just quote the main result focusing
on a $SU(2)$ subsector involving the fields (say) $Y$ and $Z$
\footnote{The labeling of the scalar fields by $Y$ and $Z$ is done
on purpose to make the connection with the coordinates used in
section 2.3 more explicit.}. There are two kind of terms in the
mixing matrix of anomalous dimension. Those corresponding to bulk
interactions in the spin chain Hamiltonian, which where already
present in the maximal case. They are nothing but the bulk terms of
a $SU(2)$ XXX spin chain. The non-maximality of the giant gives rise
to new terms proportional to the quantity
$\alpha\equiv\sqrt{1-p/N}$. The most important is a boundary term
for the spin chain Hamiltonian, coming from the correlation function
representing the exchange of a $Z$ field between the word $W$ and
the determinant
\be \langle{{{{\tilde{\cal O}}}}^{p+1 *}_W}(x)\  {\tilde{\cal
O}}^p_{W'}(0)\rangle \sim -\frac{2\lambda
\log(\Lambda|x|)}{|x|^{2\Delta_0}} \sqrt{1 -
\frac{p}{N}}\;,\;\;\;\; \text{with},\;\;\; \begin{array}{l} W =
Yw\;, \\ W' = YZw\;.\end{array} \ee
This correlation function introduces a variability in the spin
chain length at the boundaries. However, it is not a priori
clear how to deal with this $SU(2)$  spin chain of variable
length. In the following  we present how, by relabeling the
operators, we can translate the spin chain of variable length into
a more manageable bosonic lattice of fixed length and with
sources/sinks of bosons at the boundaries.

The most general word in this sector is completely specified by
the number of $Z$ fields between two consecutive $Y$ fields,
\begin{equation}
\label{label} (YZ^{n_1}Y Z^{n_2} Y \dots Y Z^{n_L} Y)_i^j \cong
|n_1, n_2, \ldots, n_L\ket\;.
\end{equation}

Thus, the word has a fixed number $L+1$ of $Y$ fields, representing
$L+1$ units of angular momentum in $Y$ direction in the string. The
number of $Z$ fields and the corresponding angular momentum is
measured by the number operator $\hat{n} = \sum_i \hat{n}_i$. The
size of the giant graviton is measured by $\hat{p} = (p + 1) \hat{I}
- \hat{n}$, where $p$ is the (fixed) total number of $Z$ fields in
the operator. The exchange of $Z$ fields occurs only at the first
and last site of the bosonic lattice. We will see that the choice of
leaving fixed the number of $Y$ fields in our labeling is in
correspondence with the gauge choice (\ref{gauge}) in the Polyakov
action.

Since there is equal probability of a $Z$ entering or leaving the
word (see appendix \ref{con}), one expects that for sufficiently large $p$
the lowest energy states have $( \bra \hat{p} \ket - p)/N \sim 0$
in the large $N$ limit. That is, the backreaction to the size of
the giant graviton should be negligible. Therefore, in what
follows we assume that $p \sim \gamma N$ with $0 < \gamma \leq 1$,
and approximate $\hat{p}/N \approx  p/N$ in all matrix elements of
the Hamiltonian. Later, we will see that this is a consistent
approximation.

The Hamiltonian dual to the anomalous dimension matrix for this
sector then takes the form
\begin{eqnarray}
\label{H}
 H &=& 2
\lambda \sum_{l=  1}^L \hat{a}_l^\dagger \hat{a}_l - \lambda
\sum_{l=1}^{L-1} (\hat{a}_l^\dagger\hat{a}_{l+1} +
\hat{a}_l \hat{a}_{l+1}^\dagger)
\nonumber \\ && +  2 \lambda \, \alpha^2
+ \lambda \,\alpha \,(\hat{a}_1^\dagger + \hat{a}_1) +
\lambda\, \alpha\,(\hat{a}_L^\dagger + \hat{a}_L)\;,
\end{eqnarray}
where $L \lesssim \sqrt{N}$ and the operators  ${\hat a}_i$ obey
the Cuntz algebra for a single species \cite{Cuntz},
\begin{eqnarray}
\label{ops} {\hat a}_i {\hat a}_i^\dagger = I\;, \quad {\hat
a}_i^\dagger {\hat a}_i = I - |0\ket\bra 0|\;,
\end{eqnarray}
where operators corresponding to different sites commute.

For our purposes it will be more useful to think of this algebra
as the $q \rightarrow 0$ limit of the deformed Weyl algebra,
$\hat{a} \hat{a}^\dagger - q \hat{a}^\dagger \hat{a} = 1$,
$[\hat{n}, \hat{a}^\dagger] = \hat{a}^\dagger$, $[\hat{n},
\hat{a}] = -\hat{a}$. The number operator $\hat{n}$ can be
constructed in terms of the oscillator operators as in
\cite{minic1}.

Terms in the first line of the Hamiltonian (\ref{H}) come from the
bulk interactions that were already present in the maximal case.
The first term indicates that each bosonic oscillator contributes
with an energy $2\lambda$ whenever its site is occupied. The
second term is a hopping interaction for bosons to move between
sites, so that the energy is reduced with bosons which are not
localized. The second line of (\ref{H}), apart from
the constant term, provides source and sink terms at the
boundaries of the bosonic lattice, which give rise to
non-diagonal boundary conditions, since the total
boson occupation number does not commute with the Hamiltonian.

In the next subsections we will discuss two different sectors of
the Hamiltonian (\ref{H}) and how they give rise to the BMN limit
of short strings and the semiclassical limit of long strings
discussed above.

\subsection{Evidence for a BMN Limit}

At this point, it could be a significant verification of the
validity of our dual description to see the BMN spectrum
(\ref{ppspec}) arising from the Hamiltonian (\ref{H}).
Unfortunately, we have not been able to diagonalize the
Hamiltonian (\ref{H}). Nevertheless, evidence for its
integrability will be given in section 5.

However, we do know the ground state:
\begin{eqnarray}
\label{ground} |\Psi_0\ket =(1-\alpha^2)^{L/2}\!\!\!\!\!\sum_{n_1,
\ldots, n_L = 0}^\infty \!\!\!\!\!\left(-\alpha\right)^{n_1 +
\cdots + n_L}|n_1,\ldots, n_L\ket\;,
\end{eqnarray}
and it has energy $E = 0$. The expectation value of the
 number operator for the ground state is,
\begin{eqnarray}
\label{groundoc} \bra\Psi_0 | \hat{n} | \Psi_0\ket =   \frac{L
N}{p} \left(1 - \frac{p}{N} \right)\;,
\end{eqnarray}
which is generically of order $L$, unless $p<<N$. Since $L
\lesssim \sqrt{N}$ we see that the backreaction to the giant is
indeed small compared to $p \sim N$. Moreover, note that setting
$\alpha = 0$ gives the familiar ferromagnetic ground state of the
maximal giant graviton:
\begin{eqnarray}
\label{groundmax}
\lim_{\alpha \rightarrow 0} | \Psi_0\ket = |0, 0,\ldots,0\ket
\simeq (YY\cdots Y)_i^j\;.
\end{eqnarray}

The expectation value (\ref{groundoc}) gives the amount
of angular momentum that the string acquires in the direction
of the movement of giant. Dividing by the fixed number of $Y$
fields we obtain the ratio of angular momentum in the two
directions of the $S^5$.
\begin{eqnarray}
\frac{J_\psi}{J_\eta} \cong \frac{\text{number of
$Z$s in $W$}}{\text{number of $Y$s in $W$}} \approx \frac{\bra\Psi_0 | \hat{n} |
\Psi_0\ket}{L} = \frac{ N}{p} \left(1 - \frac{p}{N} \right)\;.
\end{eqnarray}
This is precisely the same ratio of angular momentum components
of the null-geodesic we used to take the Penrose limit (\ref{ratio}).
Thus, as expected,  the ground state (\ref{ground}) corresponds to a
point like string traveling with the giant, and small fluctuations
around it  should correspond to the modes of the open string
in the pp-wave background.  Unfortunately we cannot solve for the
lowest energy modes of (\ref{H}) in general. However, we can
treat perturbatively  the boundary terms ($\alpha = \sqrt{1 - p/N} \leq 1$)
and try to reproduce the spectrum (\ref{ppspec}) at lowest order in $\alpha$.

Before doing this, a few comments are in order. First, let us look at the case
of the maximal giant graviton ($\alpha = 0$).  In this case there is a precise
dictionary between the plane wave excitations and the corresponding gauge theory
operators \cite{BHLN}.  The ground state (\ref{groundmax}) is excited by adding
$Z$ fields to the word $W$ with some momentum determined by the boundary conditions.
For example, the first excitation is given by a single $Z$:
\be
\label{onebosonmax}
|\psi^1_n\rangle =\sqrt{\frac{2}{L+1}}\
\sum_{l=1}^{L}\sin\left(\frac{n\pi l}{L+1}\right) | l\rangle\,
, \;\; \text{where}  \;\; | l \ket \cong (Y^l Z Y^{L+1 - l})_i^j\;.
\ee
The anomalous dimension of this operator is,
\be
\label{En}
E_n^{(0)}= 2\lambda\left[1-\cos\left(\frac{n\pi
}{L+1}\right)\right] \approx \frac{ \lambda \pi^2 n^2}{L^2} \, ,
\ee

On the other hand, the first excited state of the open string in the pp-wave background
has energy,
\begin{eqnarray}
\label{Ens}
E_{n \text{(String)}} \approx 1 + \frac{ \lambda \pi^2 n^2}{L^2} \, .
\end{eqnarray}
The factor of 1 can be interpreted from the increase of the classical dimension
of the {\it word} by the insertion of a single $Z$ field. Then, the anomalous
dimension (\ref{En}) agrees with the small correction in (\ref{Ens}).

However, for the case of the non-maximal giant graviton there is no precise notion
of the length of the word, and hence its classical dimension.  This is because $Z$
letters can enter and leave the word.  This means that we do not have a clear cut
distinction between what we call the ``string" and the ``giant".   In fact, we only
know the {\it average} length of the word to leading order in $L$:
$L + 1 + \bra n \ket  \sim {\cal O}(L)$.  Adding or subtracting a finite number of $Z$
fields to our operators will have no effect in the anomalous dimension in the planar
approximation.

In the spectrum of open strings on the pp-wave geometry we found what appears to be
two independent energy modes (\ref{ppspec}) differing by a factor of even integers.
But this is precisely the ambiguity in the classical dimension of the corresponding
operators. Therefore, we should only compare the {\it anomalous} dimensions of our
operators to the ${\cal O }(\lambda/L^2)$ corrections in the string theory spectrum.
In fact, we see that setting $\alpha = 0$ in (\ref{ppspec}) gives precisely the
anomalous dimension (\ref{En}).

We now want to calculate the lowest order correction in $\alpha$ to the energy of
the lowest BMN excitations of the maximal giant. For the sake of simplicity, we
focus on the  eigenstates with one-boson; that is, a single $Z$ on the word. We
write the Hamiltonian as
\be \label{H1}
 H = 2 \lambda \alpha^2 + H_0
+ \alpha V\,, \ee
where
\be H_0  = 2 \lambda \sum_{l=  1}^L \hat{a}_l^\dagger \hat{a}_l -
\lambda \sum_{l=1}^{L-1} (\hat{a}_l^\dagger\hat{a}_{l+1} +
\hat{a}_l \hat{a}_{l+1}^\dagger)\;,
 \ee
 and
 \be
 \label{V}
 V=
\lambda(\hat{a}_1^\dagger + \hat{a}_1 + \hat{a}_L^\dagger +
\hat{a}_L)\,. \ee
The ground state $|\psi^0\ket$ of the unperturbed Hamiltonian $H_0$
has zero occupation number and zero energy (c.f. (\ref{groundmax})). The one-boson
eigenstates and eigenenergies were given in Eqs. (\ref{onebosonmax}) and (\ref{En}).

The corrected energy is
\be
E_n= 2\lambda\alpha^2+E_n^{(0)}  + \alpha V_{nn}+
\alpha^2\sum_{k\neq n}\frac{|V_{nk}|^2}{E_n^{(0)}-E_k^{(0)}}\, + {\cal O}(\alpha^3)\,.
\ee
Matrix elements of $V$ in the unperturbed basis are non-vanishing
only for eigenstates that differ in one boson. Then, $V_{nn}$ is
zero and the first correction is order $\alpha^2$. To compute the
correction  to the one-boson eigenvalue at this perturbative order,
the only matrix  elements needed  are

\bea  \langle \psi^1_n|V|\psi^0\rangle \;\;=\;\;
\lambda\sqrt{\frac{2}{L+1}}\sin\left(\frac{n\pi}{L+1}\right)(1-(-1)^n)\;,\;\;\;
\text{and} \;\;\; \langle \psi^1_n|V|\psi^2_m\rangle \;,
 \eea
where $| \psi^2_m \ket$ are the unperturbed two-boson eigenstates
(in this case $m$ is a collective index). However, it is hard to
write down the two-boson eigenstates in closed form. One can not use
a dilute gas approximation, because this would only be valid for
small energies and we need $| \psi^2_m \ket$ for all energies.
Still, it is possible diagonalize numerically $H_0$ in the two-boson
subspace, {\it i.e.} to solve the eigenvalue problem for a
$L(L+1)/2\times{}L(L+1)/2$ matrix, for different finite values of
$L$ and use the results to compute the energy correction. Then, we
can see that as the number of sites increases, the correction not
only captures the $1/L^2$ dependence but also the proportionality
factor tends to the one predicted by the plane-wave spectrum
\be E_n\simeq \lambda\frac{\pi^2 n^2}{L^2}(1-2\alpha^2)\,.
\label{predic} \ee
 For instance, in table 1 we show the shift of the $n=1$ one-boson eigenvalue
 divided by  the predicted shift (\ref{predic}).
 \begin{table}[htdp]
\caption{Shift in $n=1$ one-boson energy.}
\begin{center}
\begin{tabular}{|c|c|}
\hline
$L$ & $-\frac{\Delta E_1}{2 \pi^2/(L+1)^2}$\\
\hline15 &1.122 \\
20 & 1.091\\
25 & 1.079\\
30 & 1.060\\ \hline
\end{tabular}
\end{center}
\label{default}
\end{table}%

If our Hamiltonian is integrable (see section 5), then maybe some
generalization of the Bethe Ansatz can be used to diagonalize it
and provide the final proof of the presence of the BMN limit.
Integrability is not strictly necessary. There are other forms of obtaining the BMN limit exactly
at strong coupling using different techniques \cite{BBPS, BCV, BCorr}, see also \cite{Rod}.
We also have uncovered some other qualitative features of the spectrum including the presence of
continuum bands by doing perturbative diagrams that cast doubt on a solutions of
the problem in terms of a Bethe Ansatz. We discuss these in section \ref{sec:unstable}.

\subsection{The Semiclassical Limit}

In general, it is possible to obtain a semiclassical sigma model
action governing the dynamics of a given spin chain. In \cite{Kruczenski}
it was shown that the sigma model obtained from the spin chain associated to
the $SU(2)$ sector of SYM coincides with the Polyakov action describing
the propagation of closed strings in $AdS_5\times S^5$, in a particular limit.
This comparison was extended to other sectors \cite{Her, Bel}
and to other realizations of the AdS/CFT correspondence
\cite{Fro, Ben, deM}.

Now we will obtain a semiclassical sigma model action that governs the
dynamics of bosonic lattice in the $L \rightarrow \infty$ limit, using the coherent
states basis for the path integral representation of the evolution
operator. This was already discussed in our previous note
\cite{bcv}. For completeness, we will use the $q$-deformed Weyl algebra and
will set $q \rightarrow 0$ when needed.  In appendix \ref{qch}, we give the
definition of the coherent states for this algebra.

One can then put a coherent state at each site of the bosonic lattice and construct the
propagator between the coherent states in the usual way (see
\cite{zhang}).  The resulting action is,
\begin{eqnarray}
\label{saction} S \!&=&\! \int dt \left(i \langle z_1\ldots z_L
|\frac{d}{dt}|z_1\ldots z_L\rangle - \langle z_1\ldots z_L |H
|z_1\ldots z_L\rangle \right)
\nn\\
\!&=&\!\int dt \left[-\sum_{i=1}^{L} f_q(r_i) \dot \phi_i
-2\lambda\left(\alpha^2
+\alpha(r_1\cos\phi_1+r_L\cos\phi_L)\right) \right.
\nn\\
 && \left. \!-2\lambda\left(\sum_{i=1}^{L}r_i^2-\sum_{i=1}^{L}r_ir_{i+1}
 \cos(\phi_i-\phi_{i+1}))\right) \right]\,,
 \end{eqnarray}
where we defined,
\begin{eqnarray}
f_q(x)= x^2 \frac{\exp_q'(x^2)}{\exp_q(x^2)}\;.
\end{eqnarray}
In the large $L$ limit,  the complex modulus $r$ and the complex
argument $\phi$ become  functions of a continuous variable $0\leq\sigma\leq
1$. We consider the action (\ref{saction}) in the limit $L\to\infty$
and $\lambda\to\infty$ but keeping $\lambda/L^2$ fixed and small.
The result is the following sigma model,
\begin{eqnarray}
S&=& -L \int dt \int_0^1 d\sigma
\left[f_q(r){\dot{\phi}}
+ \frac{\lambda}{L^2} (r'^2 + r^2 \phi'^2)\right] \nonumber \\
 &&- \left. \lambda \int dt\left[\alpha^2
\sin^2 \phi + \left(\alpha \cos \phi + r\right)^2 \right]
 \right|_{\sigma = 0} \nonumber \\
 &&- \left. \lambda \int dt\left[\alpha^2
\sin^2 \phi + \left(\alpha\cos \phi + r\right)^2 \right]
 \right|_{\sigma = 1}\;,
 \label{sm}
 \end{eqnarray}
where dots and primes refer to derivatives with respect to
$t$ and $\sigma$ respectively. If we take the limit $q \rightarrow 0$, the
function above reduces to $f_q(r) \rightarrow r^2/(1 - r^2)$ and we
see that the bulk action of (\ref{saction}) coincides with the
Polyakov action in the large momentum limit (\ref{stringaction}).
This gives a direct geometrical meaning to the fields $r$ and $\phi$
of the coherent states: as spacetime coordinates.

The classical Hamiltonian of the coherent states is
\begin{eqnarray}
\label{Hclass} \bra H\ket &=& \frac{\lambda}{L} \int_0^1 d\sigma
(r'^2 + r^2 \phi'^2) + \left. \lambda \left[\alpha^2 \sin^2 \phi +
\left(\alpha\cos \phi + r\right)^2 \right]
 \right|_{\sigma = 0} \nonumber \\
&+& \left. \lambda \left[\alpha^2\sin^2 \phi + \left(\alpha\cos
\phi + r\right)^2 \right]
 \right|_{\sigma = 1}\;.
\end{eqnarray}
One sees that the boundary terms will give rise to a large
anomalous dimension of order $\sim \lambda$ unless,
\begin{eqnarray}
r|_{\sigma = 0, 1} &=& \alpha = \sqrt{1 - \frac{p}{N}}\;, \\
\phi|_{\sigma = 0, 1} &=& \pi\;.
\end{eqnarray}
These are exactly the Dirichlet boundary conditions (\ref{bc}) on open strings on
non-maximal giant gravitons! The fact that the boundary terms give
rise to a large anomalous dimension is nothing but the statement
that moving the D-brane takes a lot of energy.

In the limit $L \rightarrow \infty$ we get the following classical equations of motion:
\begin{eqnarray}
\label{eom1} \frac{r \dot{r}}{(1 - r^2)^2} + \frac{\lambda}{L^2}
\partial_\sigma(r^2 \phi') &=&0 \;, \\
\frac{r \dot{\phi}}{(1 - r^2)^2} + \frac{\lambda}{L^2} (r \phi'^2
- r'') &=& 0\;.
\end{eqnarray}
Looking at the Hamiltonian (\ref{Hclass}) we see that the classical solutions to these
equations will have energies $\sim \lambda/L$, which represent a small
{\it multiplicative} correction to their ``bare" energies:
$L + 1 + \bra n \ket + {\cal O}(\lambda/L) \sim L(\text{const.} + {\cal O}(\lambda/L^2))$.
This is already a familiar result for closed semiclassical strings \cite{Tseytlinrev}.

The average number of bosons in the lattice is,
\begin{eqnarray}
\bra \hat{n} \ket = L \int_0^1 d\sigma \frac{r^2}{1 - r^2}\;.
\end{eqnarray}
This provides a way of measuring length of the spin chain in the
original XXX model formulation. In general, $\bra \hat{n}\ket$
is not conserved and therefore the string will oscillate in length.
Using Eq. (\ref{eom1}), this variation can be put simply as,
\begin{eqnarray}
\label{ndot}
\partial_t \bra \hat{n} \ket = 2 \frac{\lambda}{L} \left(1 -
\frac{p}{N}\right) \left(\phi'|_{\sigma = 0} - \phi'|_{\sigma = 1}
\right)\;.
\end{eqnarray}
Note however that we must ensure that $\bra \hat{n}\ket$ is
bounded so that neglecting the backreaction to the giant remains a good approximation.

Moreover, we see how the particular choice of spacetime coordinates
and world sheet gauge are encoded in the SYM side. We labeled our
states using the word $W$ only,  which is dual to the open string.
This is translated to the string side by choosing a coordinate
system for which the giant graviton is static. Finally, by labeling
the states as in (\ref{label}) we are explicitly distributing the
angular momentum in $Y$ uniformly along the string. This has very
strong implications in the AdS/CFT correspondence, because we are
seeing explicitly the reparametrization invariance of the string
worldsheet: the  gauge that makes the calculation more natural is
different than the one considered in other semiclassical setups
\cite{FT,kru}.

\section{A D-brane Instability?}\label{sec:unstable}

He have seen that the anomalous dimensions of the dual
operators of non-maximal giant gravitons with open strings excitations
is described in terms of a spin chain of variable length. We offered an alternative
description in terms of a bosonic lattice. There, the variability of the spin
chain's length was translated into the variability of the boson occupation
number. In this section, we study the possibility of having configurations
with the occupation number growing monotonically in time. More precisely,
we show evidence for the presence of continuum energy bands in the
spectrum of the Hamiltonian (\ref{H}). If a configuration has enough
energy to access the continuum,  a more general time dependence
than an oscillation is allowed for its mean occupation number.

In particular, we argue that in the large $L$ limit there is at
least one band that is accessible to classical long strings with
energies $\sim \lambda/L$. This has profound consequences for the
stability of the D-branes because the word $W$ in ($\ref{op1}$) can
be excited in such a way to absorb a large number of $Z$ fields from
the giant graviton. In the dual picture, this means that the string
becomes longer and longer absorbing more and more angular momentum
from the D-brane. A long string attached to the moving giant
graviton, could suffer centrifugal forces, consequence of a
non-geodesic movement, and if the string is long enough its tension
could be overwhelmed by these forces. Finally, the D-brane would
become small enough so that the mixing amplitude with closed strings
is no longer negligible (see appendix \ref{con}). In fact, way
before this happens, the string would have absorbed ${\cal O}(N)$
$Z$ fields so that the planar approximation used in the calculations
will be invalidated. The outcome of the instability is then
difficult to predict but can have important consequences in string
theory.

\subsection{One Site}
 We start with the Hamiltonian for $L = 1$:
\be
 H =2\lambda\left[\alpha^2 +\hat{a}^\dagger \hat{a}
  +\alpha (\hat{a}^\dagger + \hat{a})\right]\,.
\ee
The advantage of this case is that we can diagonalize $H$
explicitly for $q = 0$.  This is because $\hat {a}^\dagger$ and $\hat a$
act as shift-left and shift right operators for $q=0$. Thus, the matrix representation of $H$ looks like the
following matrix
\begin{equation}
H \sim \alpha^2 + \begin{pmatrix} 0 & \alpha& 0&0&\dots\\
\alpha&1&\alpha&0&\dots\\
0&\alpha&1&\alpha&\ddots\\
\vdots&\ddots&\ddots&\ddots&\ddots
\end{pmatrix}\,,
\end{equation}
and this is the same hamiltonian for a particle hopping on a semi-infinite lattice, as well as the
quadratic form that one considers when studying the potential energy of a semi-infinite collection of beads
attached by springs.

The complete set of eigenstates consists
of the ground state (\ref{ground}) with $L = 1$, which is considered as a bound state for the hopping particle
to scatter of the boundary and the continuum,
\be
|\Psi(k)\rangle  = \sum_{n=0}^{\infty}\left[\sin kn +
\alpha\sin k(n+1)\right]|n\rangle\, ,\; \;\;\;{\rm with\ } 0\leq
k\leq\pi\,,
\ee
and energy
\be E(k)= 2\lambda(1 +2 \alpha\cos k+\alpha^2)\,.
\label{b1}
\ee

The gap with the ground state turns out to be
\begin{equation}
2\lambda(1-\alpha)^2\;.\label{eq:gap}
\end{equation}
Moreover, the maximal energy of the
continuum is $2\lambda(1+\alpha)^2$. These continuum states follow
a delta-function normalization
\be \langle\Psi(k')|\Psi(k)\rangle = \frac{\pi}{2}(1+2 \alpha\cos
k+\alpha^2) \delta(k'-k)\,. \ee
We can then use them to build normalizable wave-packets
\be |\phi\rangle =  \int_0^\pi dk f(k) |\Psi(k)\rangle\, ,
\label{fi} \ee
choosing a function $f(k)$ so that $\langle\phi|\phi\rangle = 1$
and the initial occupation number $\langle\phi|\hat n(0)
|\phi\rangle$ is  finite. Then we can have normalized states with
finite energy but with growing occupation number.

For instance, the wave-packet
\be\label{wp} |\phi\rangle \propto  \int_0^\pi dk \sin k  |\Psi(k)\rangle =
\frac{\pi}{2} \left( \alpha | 0 \ket + | 1\ket \right)\,, \ee
increases its mean occupation number while it evolves, as it is
depicted in Figure \ref{fig:packet.eps}.

\myfig{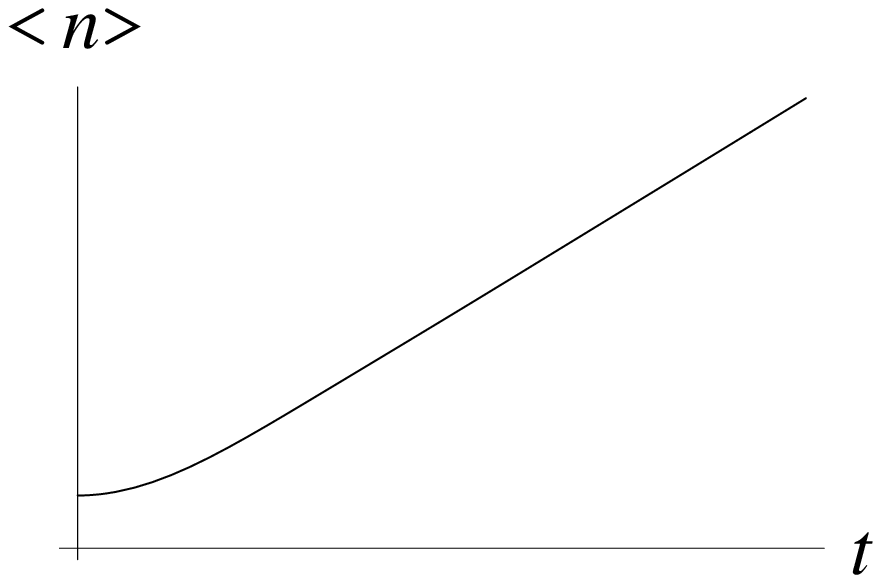}{4.5}{Mean occupation of wave-packet (4.8).}

This continuum spectrum feature is not only a property of $q=0$. Indeed, when $q<1$ we always find the same
behavior. This is because for sufficiently large $n$ (occupation number), the hamiltonian asymptotes exponentially
quickly to a similar form

\begin{equation}
H \sim \begin{pmatrix} \ddots & \ddots& \ddots&\ddots&\ddots&\ddots\\
\ddots&\alpha&1&\alpha&\ddots&\ddots\\
\ddots&\ddots&\alpha&1&\alpha&\ddots\\
\ddots&\ddots&\ddots&\ddots&\ddots&\ddots
\end{pmatrix} \, ,
\end{equation}
since $q^n\to 0$.

\subsection{Two Sites}

For two sites we can no longer solve  for the exact spectrum but
we can do (degenerate) perturbation theory around the maximal case
($\alpha = 0$ and $q=0$). The unperturbed Hamiltonian for two sites is,
\begin{eqnarray}
H_0 = 2 \lambda\, \hat{a}_1^\dagger \hat{a}_1 + 2 \lambda \,
\hat{a}_2^\dagger \hat{a}_2 - \lambda (  \hat{a}_1^\dagger
\hat{a}_2 + \hat{a}_1 \hat{a}_2^\dagger )\;.
\end{eqnarray}

Since this Hamiltonian correspond to an integrable spin chain, we
can diagonalize it using the usual Bethe Ansatz. The eigenstates
are (e.g. see \cite{sam}),
\begin{eqnarray}
\label{states} |\Psi^{(0)}_n(k)\ket = A_n(k) \sum_{l = 0}^n
\left[2 \sin(k \,l) - \sin (k (l+1)) \right] |l,n\ket\;,
\end{eqnarray}
where  the states $|n_1 ,n_2\ket$ have been re-labeled $|n_1, n_1
+ n_2\ket \equiv |l,n\ket$ to reflect the fact that the number
operator commutes with $H_0$. The energy of these eigenstates is
given by,
\begin{eqnarray}
\label{Ek} E(k) = 2 \lambda (2 - \cos k)\;,
\end{eqnarray}
and the complex momentum $k$ is determined by the equation,
\begin{eqnarray}
\label{roots}
 4 \sin(k n) - 4 \sin (k (n+1)) + \sin(k (n+2))
= 0\;.
\end{eqnarray}

We now want to consider first order perturbation theory in
$\alpha$ around these eigenstates. Again, the perturbing potential
is (\ref{V}) and we ignore the constant term in (\ref{H1}) which
is  $O(\alpha^2)$. In particular we are interested in the
continuum energy bands that could arise from turning on the
perturbation. For that to occur we will need to have an infinite
degeneracy or an accumulation point in $k$ in the large $n$ limit.

Let us start by looking at the infinite degenerate roots of
(\ref{roots}).  It is easy to show that for a given $n$, the
degenerate roots have the form $k = a\pi/b$ with $a, b \in
\mathbb{Z}$ and where the degeneracy occurs for states
$|\Psi_{n'}^{(0)}\ket$ with $n' = n + \mathbb{Z} b$. Now, the
perturbation (\ref{V}) has non-zero matrix elements only between
unperturbed states differing in $n$ by one. Therefore, at first
order in perturbation theory we only need to consider the
degenerate roots: $k = 0 , \pi$ (we can take $k \in [0,\pi]$
without loss of generality).

For $k = 0$, the unperturbed energy is $E^{(0)}= 2 \lambda$. To
find the first order correction in $\alpha$ we need to diagonalize
$V$ in the subspace of the unperturbed energy eigenstates
(\ref{states}). However for our purposes, we only need to look at
the large $n$ limit to see the presence of the continuum. One can
show that in this limit,
\begin{eqnarray}
V |\Psi_n^{(0)}\ket \sim 2 \lambda \alpha \left(
|\Psi_{n+1}^{(0)}\ket +  |\Psi_{n-1}^{(0)}\ket \right)\;.
\end{eqnarray}
Thus the asymptotic form of the first order correction to the
eigenstates is going to be of the form of a superposition of plane
waves $|\Psi_n^{(1)}\ket \sim \sum_n e^{i p n} |\Psi_n^{(0)}\ket$
with continuum momentum $p$ and energy:
\begin{eqnarray}
E^{(1)} = 4 \alpha  \lambda \cos p\;.
\end{eqnarray}

If we follow the same procedure with the root $k = \pi$ we find
that at large $n$ the matrix elements of $V$ vanish and so there
is no continuum band at this energy. Next, we consider the case of
complex $k$ and we find an exponential accumulation point at $k =
i \log 2$ with energy $E^{(0)} = 3 \lambda/2$. The asymptotic form
of the action of the perturbation is,
\begin{eqnarray}
V |\Psi_n^{(0)}\ket \sim \frac{3}{2} \lambda \alpha \left(
|\Psi_{n+1}^{(0)}\ket +  |\Psi_{n-1}^{(0)}\ket \right)\;,
\end{eqnarray}
and so we get a continuum band with,
\begin{eqnarray}
E^{(1)} = 3 \alpha  \lambda \cos p\;.
\label{b2}
\end{eqnarray}

\subsection{Multiple Sites}

For multiple sites a quantum analysis is no longer feasible.
Nevertheless the presence of continuum bands can be seen directly
from a classical analysis. For any number of sites the classical
limit is $\hbar \rightarrow 0$ and thus we can use the coherent
state Hamiltonian. For completeness, we will work with $0 \leq q
\leq 1$ and then take the limit $q \rightarrow 0$ when needed.

Looking at the action (\ref{saction}) we see
that our system is subject to the constraints,
\bea
p_{r_i}&=&0\, ,\nn\\
p_{\phi_i}+ f_q(r_i)&=&0\, , \eea and then, the phase space is
isomorphic to the configuration space $(r_i,\phi_i)$, which is a
product of $L$ discs $D_q$ of radii $\frac{1}{1-q}$.
Using the auxiliary constants $r_0=r_{L+1}=\alpha$ and
$\phi_0=\phi_{L+1}=\pi$ the classical Hamiltonian can be simply
posed as
\be {\cal H}_L= 2\lambda\sum_{i=0}^{L} r_i\left(r_{i}
-r_{i+1}\cos(\phi_i-\phi_{i+1})\right)\, . \label{HL} \ee
In particular, let us first consider the one-site classical
Hamiltonian
\be {\cal H}_1= 2\lambda(\alpha^2 + r^2 - 4 \alpha r \cos\phi )\,
. \ee
\myfig{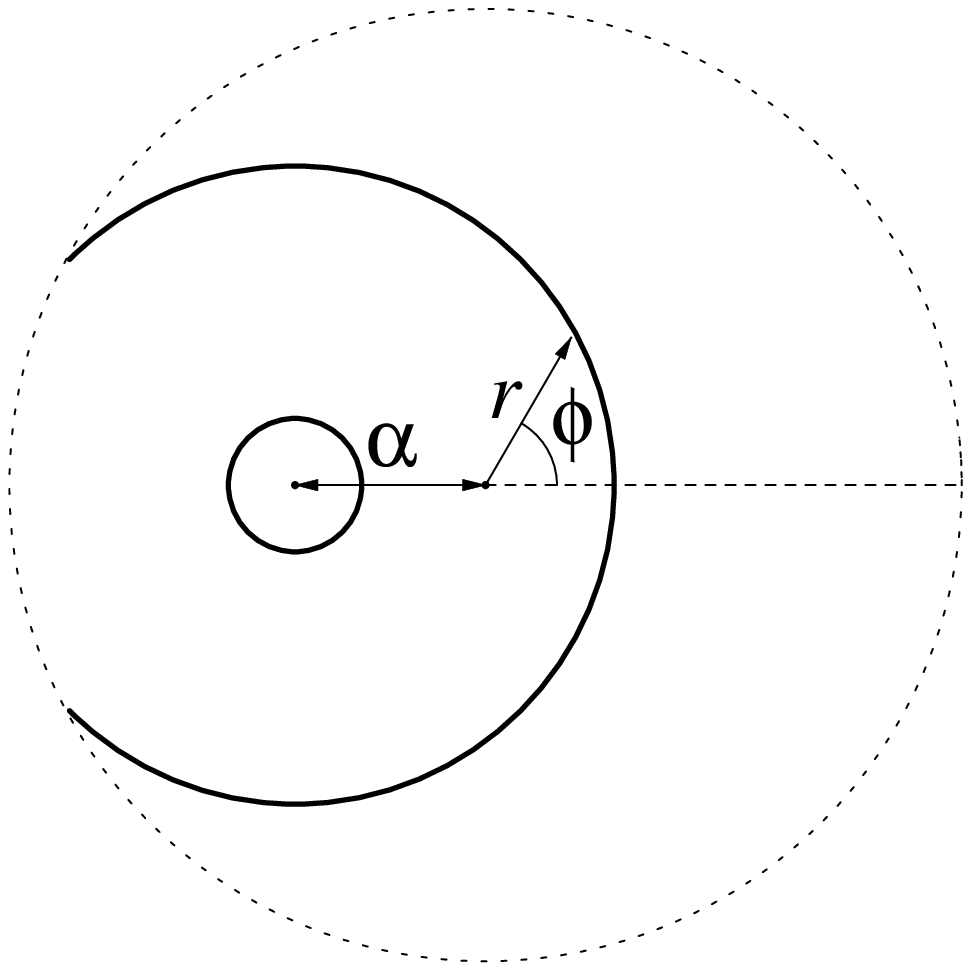}{4}{Closed and open orbits in $D_q$.}

Curves of constant energy, which is the unique conserved quantity,
determines the trajectories in the two-dimensional phase space
$D_q$. In the present coordinates, the ${\cal H}_1=E$ curves are
nothing but circles whose centers are displaced a distance
$\alpha$ from the center of the disc $D_q$ (see figure
\ref{fig:orbits.eps}). The radii of these orbits are given by the
energy through $\sqrt{E/2\lambda}$. As long as
$E<2\lambda(\frac{1}{1-q}-\alpha)^2$ the orbits are closed
circles. However, for energies
$2\lambda(\frac{1}{1-q}-\alpha)^2<E<2\lambda(\frac{2}{1-q}-\alpha)^2$
the orbits are open arcs. In a semiclassical approach, energies
for which the trajectories are open in the phase space give rise to a
the continuum in the quantized system. The proper measure of the continuum is the
phase space area between energies $E$ and $E+\delta E$. This area has
to become infinite once the trajectories reach the boundary (otherwise one
would find that the phase space has finite area, and then the total number
of states would have to be finite).

 The minimal and maximal
energies of the continuum in the $q\to 0$ limit  are seen to agree with
the results of the quantum perturbation theory.

We now finally come to the case of arbitrary number of sites. In
this case we will only calculate the energy gap between the ground
state and the first continuum band. To obtain the energy for which
the continuum begins,  one has to look for the minimal energy for
which the energy constant hypersurfaces intersect the boundary of
the phase space $D_q^L$. To this end one needs to compute and
compare the absolute minimum of the resulting $L$ functions when
one of the $r_i$ is set to $\frac{1}{1-q}$. These $L$ functions
are bounded from below by quadratic polynomials on the radii $r_i$
\be {\cal H}_L^j=\left.{\cal H}_L\right|_{r_j=\frac{1}{1-q}} \geq
{\cal M}_L^j=2\lambda\sum_{i=0}^{L} r_i\left.\left(r_{i}
-r_{i+1}\right)\right|_{r_j=\frac{1}{1-q}}\, , \label{ML} \ee
and coincide with them when all angles $\phi_i$ are taken to be
$\pi$. Then it suffices to find out the minima of the ${\cal
M}_L^j$. The stationary points should satisfy in all cases the
recursive equation
\be \label{recu} 2r_i - r_{i+1} -r_{i-1}=0\,, \ee
subject to the boundary conditions
\bea \label{boco}
r_0\! = \!\alpha\, ,\;\;\; r_j\! = \!\frac{1}{1-q}\, ,\;\;\;
r_{L+1}\!=\!\alpha\, . \eea
The recursive equation (\ref{recu}) is fulfilled by a linear
dependence of  $r_i$ on the site labeling $i$, with the constants
adjusted to also fulfill the boundary conditions (\ref{boco}). The
critical point of each ${\cal M}_L^j$ is
\bea r_i^*\!&=&\!\left(\frac{1-\alpha(1-q)}{j(1-q)}\right)i+\alpha
\;\;\;\;\;\; \;\;\;\;\;\; \;\;\;\;\;\; \;\;\;\;\;\;  \;\;\;\;\;\;
\;\;\;\;\;\; \;\;\;\; {\rm if}\; i \leq j\ , \nn
\\
r_i^*\!&=&\!\left(
\frac{\alpha(1-q)-1}{(L+1-j)(1-q)}\right)i+\frac{L+1
-j\alpha(1-q)}{(L+1-j)(1-q)}\;\;\;\;\; {\rm if}\; i > j\ . \eea
It is easy to see that each critical point is a minimum of each
${\cal M}_L^j$,
 since all the eigenvalues of the corresponding Hessians are strictly positive. The
 energy evaluated in these points is
\be \left.{\cal
H}_L^j\right|_{r_i=r_i^*,\phi_i=\pi}=\frac{\lambda(L+1)(1-\alpha(1-q))^2}{j(L+1-j)(1-q)^2}\,
. \ee
As a function  of $j$, that takes values $1\leq j\leq L$, the
minimal energy is obtained for the $j$ corresponding to the
central site. So, the minimal energy for which constant energy
hypersurfaces intersect the phase space boundary is
\be E_{\rm cont}= \left\{
\begin{array}{cll}
  \frac{4\lambda(1-\alpha(1-q))^2}{(L+1)(1-q)^2}&   &{\rm if}\ L\ {\rm is\ odd}   \\
  & & \\
  \frac{4\lambda(L+1)(1-\alpha(1-q))^2}{L(L+2)(1-q)^2} &   &{\rm if}\ L\ {\rm is\ even}
\end{array}
\right. \,.\ee
Let us now turn back to case $q\to 0$. For $L=1$ we have again
$E_{\rm cont}=2\lambda(1-\alpha)^2$, which is the minimum of the unique
band (\ref{b1}) of the quantum system. For $L=2$, we have
$E_{\rm cont}=3\lambda/2(1-\alpha)^2$ which is the minimum of the lowest
band (\ref{b2}) computed perturbatively for small $\alpha$. We are
eventually interested in the limit of a large number of
sites $L$. According to the classical analysis, the minimum of the lowest band is,
in that case,
\be E_{\rm cont}= \frac{4\lambda(1-\alpha)^2}{L}\, . \ee
As we argued, from this  energy the quantized system will have a
continuum spectrum. Then,  this is the amount of energy needed for
having configurations with a  mean occupation number growing in
time. Note that this energy is accessible to long semiclassical
strings.

Let us end this section, presenting how
the transition from oscillatory  to non-oscillatory solutions near
the value $E_{\rm cont}$ is observed in  numerical solutions.
In figures \ref{fig:array1.eps},
\ref{fig:array2.eps} and \ref{fig:array3.eps} we show different classical
solutions of the system with $L=30$ and their corresponding
occupation numbers. The 3 cases were solved with
similar initial data. Only the amplitude of the initial profiles ($r_i$, $\phi_i$)
were slightly varied to  change the energies.  Moreover we chose the energies
to be near the critical value $E_{\rm cont}$. For energies below
$E_{\rm cont}$, no $r_i$ can tend to 1, so the mean occupation
number is kept bounded. But as long as the energy is bigger than
$E_{\rm cont}$ it is possible to find solutions with some $r_i$
approaching 1 and a mean occupation number growing monotonically
in time.

\myfig{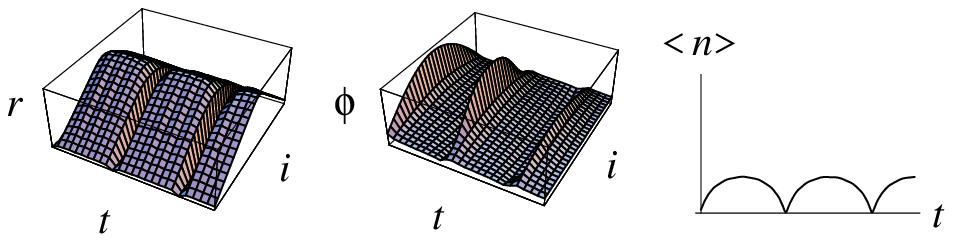}{10}{$L=30$ solution with energy $E=0.94 E_{\rm
cont}$.}

\myfig{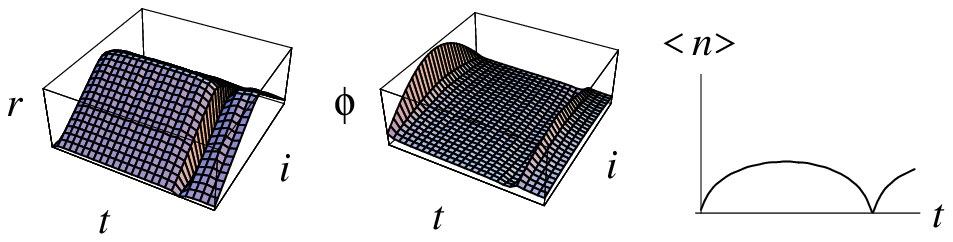}{10}{$L=30$ solution with energy $E=0.99 E_{\rm
cont}$.}

\myfig{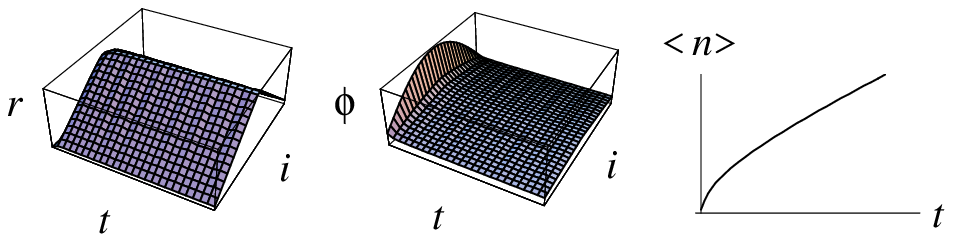}{10}{$L=30$ solution with energy $E=1.05 E_{\rm
cont}$.}

\subsection{A toy model for the D-brane instability}

We have found, surprisingly, that the spin chain model associated to a string attached to a non-maximal giant
exhibits a continuous spectrum. Naively, this is counter to our intuition from AdS, as we have a finite energy
state, and in the dual field theory, there are only finitely many states with energy less than $E$. This means
that the true spectrum of the string plus D-brane system ultimately must be discrete.

Considering how we got our spin chain model, we have to notice that we had various assumptions built in our
computations. First, the giant has energy of order $N$, while the string should have an energy of order
$\sqrt N$, roughly, this  corresponds to occupation numbers lower than $\sqrt N$. This is required for the
planar approximation to Feynman diagrams to make sense. If the total angular momentum of the spin chain becomes
large with respect to $\sqrt N$, the planar approximation breaks down. This means that the Hamiltonian in terms
of the boson spin chain we have described is only valid in some regime (where the occupation numbers are not
too large), and when we get out of that regime, the fact that $N$ is finite is important. As we take $N$ large,
the bosonic spin chain model realm of applicability grows. So the continuum spectrum and bands we have been
discussing are a property of the strict large $N$ limit. At finite $N$, the continuum spectrum is resolved into
a discrete spectrum with an extremely fine spacing of the eigenvalues. This would be very hard for an observer to
determine. This continuum spectrum, which is associated with large quantum numbers for the occupation numbers of
the chain  has to be interpreted in terms of classical physics (this is the same reasoning that went into the work
\cite{GKP2}). At large quantum numbers, the problem becomes classical, and we should be able to come up with a
classical argument that shows that the strings should grow in size, so long as string interactions and the brane
back reaction can be ignored (these are the $1/N$ effects that the planar approximation does away with).

Now, we want to come up with a model of the system that clearly shows the instability, without the complications
of the $AdS_5\times S^5$ geometry. The idea is simple: we have a charged D-brane in the presence of a magnetic RR
background.  Because of the background, a moving D-brane is accelerated with respect to geodesic free-fall by a
Lorentz force. Since fundamental strings are not charged with respect to the RR-background, they should prefer to
follow geodesics.

This motivates the following toy model for the giant graviton setup: the ends of the strings are attached to the
D-brane, so they are accelerating. By the equivalence principle, we can think of the same system as a string with
fixed ends suspended in a gravitational field. A close analogy of the setup is the act of  suspending cables
between telephone poles. The telephone poles are replaced by  the D-brane itself , and it is supported in place
by the RR-background. The cable between the telephone posts is the string. Unlike conventional cables, the string
tension is constant (a fundamental constant in perturbative string theory). This means that in principle the
string can weigh more that the tension of the string can support. This requires a long string (long enough so
that the weight of string between the two posts is bigger than the string tension), whose length depends on the
strength of the gravitational field.

Since the string can not break (we are assuming that we are in the strict planar approximation), the string will
grow by falling,  if it is long enough to begin with (this can be accomplished by separating the putative
telephone posts sufficiently). In fact, the string will keep on falling forever if string interactions are turned
off. Otherwise, the string will get so long that eventually there is a finite probability per unit time that the
one long string will break into a closed loop plus a shorter string suspended between the post (this can be
understood as the emission of gravitational waves, or other fields, by an accelerated object).

Our toy model clearly exhibits the instability of the string system that we found with our spin chain model.
It also provides for the presence of the gap: if the telephone posts are close together, the minimal energy
string will be stable. Indeed, many small perturbations of the configuration will be allowed. To make the string
fall, we might need to stretch it a lot, but once we have enough weight in the system, the string can not sustain the
total weight of  the string, and the instability we discuss will have set in. Because we have a regime where the
string is stable, this regime of lower energy will have a discrete spectrum, and when we reach the instability, we
will get a continuous spectrum. This gap is determined by the distance between the telephone posts (this can be
related to the total angular momentum of the string ground state), and by the effective gravitational field. This
effective gravitational field is determined by how much is the giant graviton accelerating: the bigger giants move
slowly, so their effective gravity is small, while small giants move fast and accelerate a lot more.

This means that the gap in the spectrum should become smaller as the D-brane becomes
smaller (in our spin chain notation, $\alpha$ becomes bigger). This is exactly what we see from eq.
\ref{eq:gap}, as well as from understanding the details of figure \ref{fig:orbits.eps}.

\section{Evidence for Integrability}
\label{sec:integrability}

In this section we present numerical evidence for the
integrability of the classical Hamiltonian (\ref{HL}) for $q \in
[0,1]$. For
 $L$ sites, this would correspond to have $L$ constants of
motion\footnote{Moreover the constants of motion should be
compatible with each other, which means that the Poisson brackets
with each other should vanish.}. Then, trajectories in the
$2L$-dimensional phase space take place in a $L$-dimensional torus.
However, with the exception of separable systems, there is no
systematic procedure for finding constants of motion. Even in the
rather simple case of two sites, we do not know another constant of
motion apart from the energy whenever $\alpha\neq 0$. Hypersurfaces
of constant energy are 3-spheres in $D_q^2$, and if a second
constant of motion existed, the motion would be in a 2-torus
included in the 3-sphere. Then, the intersection of this 2-torus
with a any hypersurface (of dimension 3)  would be in general a
closed curve in the phase space. We can verify if this is the case
by studying numerical solutions of the system. For example, in
Figure \ref{fig:points.eps} we plot the values of coordinates
$(r_2,\phi_1,\phi_1)$ of a given solution for many different times
when the variable $r_1$ takes the value $1/2$. The fact that points
lie in a closed curve is a strong indication that the motion is
taking place in a 2-torus and the system is integrable. One can try
this for many different hypersurfaces and one always gets closed
curves. Similar evidence is found for cases with higher number of sites.

\myfig{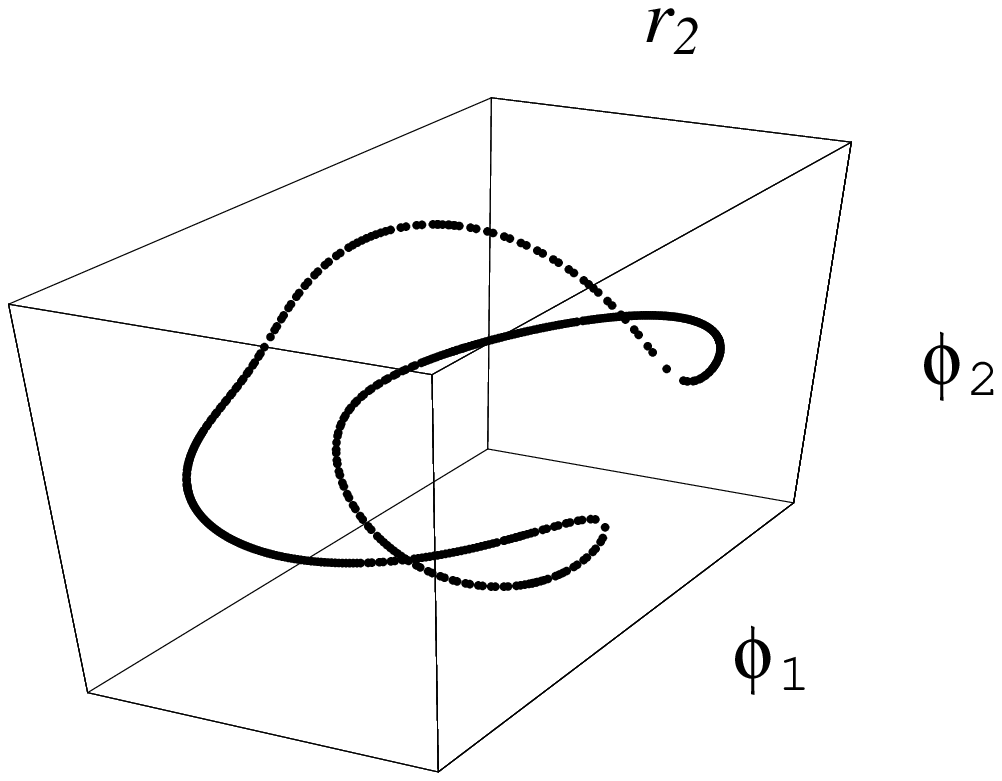}{4.5}{Intersections of the a solution with $r_1=1/2$.}

Note that for $q = 1$ the deformed operators (\ref{defops}) become
the usual harmonic oscillators and the system becomes a lattice of
ordinary bosons. In this case quantum and classical integrability
follow trivially. For $q \neq 1$ the classical Hamiltonian
(\ref{HL}) can be seen as a deformation of the Hamiltonians
studied in \cite{Calogero, K1, K2}. We now conjecture that our
family of Hamiltonians parameterized by $q, \alpha$ is indeed a two
parameter family of (quantum) integrable Hamiltonians.

Because the spectrum of the quantum hamiltonian has a continuum, it is very unlikely that the system can be solved
via a simple Bethe Ansatz for all $\alpha$, even though for $\alpha=0$ the system can be solved via a Bethe
Ansatz. The distinction between $\alpha=0$ and $\alpha\neq 0$ resides in the fact that the constant energy surfaces
are compact in the first case, while they are non-compact in the second case (they intersect the boundary). This
non-compactness of the energy surfaces for the spin model upsets the usual Bethe Ansatz intuition for finite
systems. In finite systems the Bethe ansatz hinges in having some discrete conserved number of quasi-particles,
and this produces a system of algebraic equations that describe scattering of these quasi-particles with each
other and the boundary. The solutions  of these Bethe equations give a finite number of roots that one identifies
with the discrete spectrum of the finite system.

To have a continuous spectrum one needs one of two conditions: either the spin chain is infinite, or the conserved
``number of quasiparticles" is described by a continuous parameter. The first case does not describe our system
well, as we have found that the ground state has a well defined finite length (average number of spins). The
second case would not  produce algebraic equations that solve for the spectrum and we would be hard pressed to call
such a system a Bethe Ansatz.

However we seem to have an integrable system with a continuous spectrum. We should contrast this observation with
the observation of \cite{Agarwal} that the Bethe Ansatz seems to break  down at two loops for the maximal giant
graviton, or that the BMN limit breaks down \cite{OY}. There are also other hints  that the asymptotic Bethe Ansatz conjecture is incomplete \cite{BCV}. We interpret all of these 
facts not as a breakdown of integrability, but instead as a breakdown of applicability of a Bethe Ansatz.

\section{Discussion}
\label{sec:discussion}

In this paper we have shown an example that dynamical finite
D-branes can be treated consistently in the AdS/CFT setup. In
particular we showed that we were able to characterize the D-brane
in the most conservative form: as a geometric locus where open
strings can end. We saw that this property could be derived exactly
from the dual field theory.  We were able to do this not just for a
maximal giant graviton, but also for smaller giant gravitons that
move in the AdS spacetime. The calculations required in field theory
were rather subtle and we found a lot of surprises in trying to
explain the dynamics of this system in detail.

The first surprise we found was that the trajectory of the spherical
giants we studied contained families of null geodesics, and that it
was possible to take a standard Penrose limit along these
trajectories. We found that this limit gave us a different
coordinate system than the usual one, because it was adapted to the
D-brane being static.

We then found that is was possible in the limit of large quantum
numbers to get the classical  ratio of different components of
angular momentum carried by the ground state of the string. Part of
it was along the direction of angular momentum of the giant, and the
string and giant in principle exchange angular momentum between
them.

We then looked for a description of these semiclassical states of
the AdS geometry directly in the dual field theory in the
perturbative weak coupling approximation. We found a spin chain
model with variable numbers of sites at the boundary. We were also
able to describe this system in terms of a boson chain model, where
the individual bosons obeyed Cuntz algebra oscillator relations. The
Hamiltonian we found was quadratic in generalized raising and
lowering operators. It results as a limit $q\to 0$ of a $q$-deformed
chain of harmonic oscillators with nearest neighbor interactions.
The boson system did not preserve the total boson occupation number,
so the boundary conditions we found are of non-diagonal type.

A big surprise was that when we tried to diagonalize the boson chain
model we found a continuous spectrum of the effective hamiltonian.
This was very unexpected when we started the project, but we found
an analogy that made it intuitively obvious and necessary for the
AdS/CFT correspondence to work. The analogy consists on thinking of
the system as an open string whose ends are accelerated because they
are attached to a D-brane,  in the presence of a RR background. The
background accelerates the D-brane, and the D-brane drags the string
itself. The continuous spectrum was argued to be related to an
instability that appears only if the string is too long and can not
support its own weight. This instability might be interesting for
the study of cosmic strings, because it can serve as a mechanism to
seed long strings in an expanding universe.

We also found numerical evidence that the open string attached to
the giant graviton was in general to one loop order governed by an
integrable system. Our evidence is numerical and classical. We also
found that the continuous spectrum seems to prevent the
integrability of the system to be described by a Bethe Ansatz.

Nevertheless, despite this progress, there are
many loose ends and open problems left to study.  Here we would like to
list some of them to encourage future research along these lines.

\begin{enumerate}
\item{\bf The Plane Wave Spectrum and Integrability}

The issue of the integrability of the Hamiltonian (\ref{H}) is important to prove the presence of the BMN
spectrum. However, the fact that the spectrum has continuum energy bands suggests that integrability might
be realized in a way quite different from the usual Heisenberg spin chain. In fact, since we have absorption
and emission from the boundaries, one wonders if the usual formalism using the boundary reflection matrices
could be generalized to take into account this effect \cite{int}. Whatever the answer, this problem can have
implications beyond the study of the AdS/CFT correspondence and should be very interesting to resolve.

\item{{\bf The $D$-brane Instability}}

As we mentioned before, the end state of the unstable giant graviton is difficult to predict from our
current results. The reason is that all the approximations used in our field theory calculations are invalidated
when the word's length describing the string is large with respect to $\sqrt N$. In any case, the end state of this instability can be important
for the study of certain classical spacetimes. In particular we know that the 1/2 BPS supergravity solutions
found in \cite{LLM} can be interpreted as coherent states of these giant gravitons (and the AdS giants). One
can excite massive open string modes on these backgrounds and ask what effect  the instability studied here
can have on these backgrounds (one might imagine that these might serve as a mechanism to
transport charge between droplets for example). This can be important for the stability of black holes since one  can imagine that
these will be constructed by exciting many open string modes on the D-branes. Finally, one can ask if this
instability is also present in the case of AdS giant gravitons.

\item{\bf Beyond the $SU(2)$ Sector}

In this paper we only considered open strings with  two angular momenta on the sphere. In other words, an
$SU(2)$ subsector of the corresponding gauge theory operators. However, it is interesting to go beyond this
sector. At this moment we do not know how to write down a ``nice" Hamiltonian describing the full $SO(6)$
excitations on the sphere.  However,  one can consider an $SU(3)$ subsector by using the three holomorphic
fields: $X$, $Y$ and $Z$. The generalization of our Hamiltonian is straightforward. The most general word
can be labeled as,
\begin{eqnarray}
\label{su3word} (YZ^{n_1} X Z^{n_2} Y \cdots Z^{n_L} X)_i^j \cong |
s_1, n_1,s_2,n_2, \ldots, n_L, s_{L+1}\ket\;,
\end{eqnarray}
where $s_i =\uparrow\text{or} \downarrow \, \cong  X\, \text{or}\, Y$, is an $SU(2)$ spin label and $n_i$
is the usual bosonic occupation number.

The interactions between the sites are much the  same as before except when any  $n_i = 0$. In this case we
have the additional permutation interaction between the two spins $s_{i}$ and $s_{i+1}$ (see Eq. (\ref{bulkH})).
One can then write the Hamiltonian as,
\begin{eqnarray}
H = H_{\text{boson}} + \lambda \sum_{l = 1}^{L}  (1 -
\hat{P}_{l,l+1}) (1 - \hat{a}_l^\dagger \hat{a}_l)\;,
\label{s3Hami}
\end{eqnarray}
where $H_\text{boson}$ is the Hamiltonian for  the bosonic lattice (\ref{H}), and
$\hat{P}_{l,l+1} = \frac{1}{2}(1 + 4 \vec{S}_l \cdot \vec{S}_{l+1})$ is the permutation operator acting
on the spin sites.

Constructing the action for the coherent  states goes as usual (see appendix \ref{su3app}). We get,
\begin{eqnarray}
\label{Sthree} S &=& -L \int dt \int_0^1 d\sigma \left[\frac{r^2}{1-
r^2}\, \dot{\phi} + \frac{1}{2} \cos\theta \, \dot{\varphi}  +
\frac{\lambda}{L^2} (r'^2 + r^2 \phi'^2) \right. \nonumber
\\
&&+ \left. \frac{\lambda}{4 L^2} (1 - r^2) (\theta'^2 +
\sin^2\theta\, \varphi'^2)\right]\;,
\end{eqnarray}
where the angles $\theta, \varphi$ label the $SU(2)$ coherent states. On the other hand, one can study the
dual string theory and write the Polyakov action for open strings as we did in section \ref{OSGG}. Using the
natural coordinates adapted to the brane, and making the expansion in large total angular momentum in the $X$
and $Y$ directions we get the action (\ref{Sthree}) with the exception that the last term has a different $r$
dependence: $\sim\int dt d\sigma (1 - r^2)^2 (\theta'^2+\sin^2\theta\,\varphi'^2)$. It would be interesting
to understand the nature of this discrepancy.

\item{\bf Multiple D-branes}

Recently it was shown how to write down the operators corresponding
to multiple giant gravitons with strings attached to them
\cite{david}. In general the combinatorics are hopelessly
complicated. However one can study the case of two giant gravitons
with two strings stretching between them (we need two strings to
satisfy the Gauss constraint \cite{david}). Intuitively one would
expect Hamiltonians like (\ref{H}) but with more general
coefficients for the boundary terms (different values at each
boundary). In this case there can be a ``current" of $Z$ fields from
one giant to the other. It would be interesting to see what happens
in this case.  We expect that the giants attract each other. We
would also expect the same instability we found for these models.
Moreover, one should be able to measure the distance between the two
giants using the sigma model representation of the coherent states.
More generally, one should also be able to study D-branes at angles,
and measure the angles between them by finding the spectrum of open
strings stretching between them.

For many giants the easiest description seems to be in terms of the matrix model of \cite{BBPS}. However it is
still not clear how to introduce the open string excitations in that language\footnote{See \cite{bcv} for the
emergence of closed string BMN excitations in the matrix model}. It would be interesting to re-derive the results
presented in this paper using the matrix model formalism. This can also give a better understanding of the
backreaction of the giant gravitons with massive open string modes.

\item{\bf AdS Giants}

One can try to extend the results of this paper to the study of the giant gravitons that expand in the AdS space.
The operators are similar to the ones considered here but with the $\epsilon$ symbol replaced with totally
symmetric tensor contractions.
 The combinatorics can be done in a similar way but the field theory interactions are more
complicated when we consider an open string spinning along the AdS directions. This is because, instead of scalar ``letters"
we need to use covariant derivatives. However, one expects a simple description in terms of spin chains (or a
bosonic lattice) as for closed strings \cite{Beisert1, BS}.

\item{\bf Giants of critical angular momentum}

In this paper we have considered giant gravitons whose angular momentum grows proportionally to $N$. It would
be interesting to study gravitons with angular momentum $p \sim \sqrt N$. For  $p^2/N$ fixed and $p^2/N \ll 1$
the gravitons should be described as point-like supergravity modes (as it is done in the BMN limit). However, for
$p^2/N$ fixed and $p^2/N \gg1$ the appropriate description is in terms of expanded D3-branes. From (\ref{p})
one immediately realizes that the angle $\theta_0$ goes now to zero in the large $N$ limit and each element of
the giant is traveling in an almost null geodesic. Moreover, from (\ref{rgg}) it is evident that the radius of
the giant remains finite in the limit. Then, a short open string attached to this giant is effectively described
by an open string attached to a spherical D3-brane of finite radius in a pp-wave background \cite{taka}. The
string does not need to be spinning along the giant. Then, bosonic lattices with a finite and small number of
sites should be used to describe the dilation operator in the dual field theory.

\end{enumerate}

\section*{Acknowledgements}
D.B. would like to thank A. Ludwig for conversations related to this work.
D.B.and S.E.V.  work was supported in part by a DOE
award, under grant DE-FG02-91ER40618. D.H.C. work was supported by Fondecyt grant 3060009.  Institutional grants
to CECS of the Millennium Science Initiative, Fundaci\'on Andes, and
support by Empresas CMPC are also acknowledged.
S.E.V. work was supported in part by an NSF graduate fellowship. 

\appendix
\section{Penrose limit}
\label{apl}

The Penrose limit is accomplished defining new coordinates by means
of a linear transformation and taking an appropriate $R\to\infty$
limit. The coordinate playing the role of the curve parameter,  must
appear with the same coefficient in $t$, $\psi$ and $\eta$, if we
want to capture the geometry near the null trajectory (\ref{null}).
The rest of the coefficients can be fixed by demanding the $R\to
\infty$ limit to be well-defined. Then, we consider the
transformation
\bea \label{coor} && t = u +\frac{v}{R^2} \, , ~  ~ ~ ~ ~ ~ ~ ~ ~ ~
~ ~ ~ ~ ~ ~ ~ ~ ~ ~ ~ ~ ~ ~ ~ ~
\!\rho= \frac{r}{R} \nn\\
&& \psi = u -\frac{v}{R^2} +\tan\theta_0 \frac{x}{R}  \, , ~  ~ ~ ~
~ ~ ~ ~ ~
\theta= \theta_0+\frac{y}{R} \nn\\
&& \eta =u -\frac{v}{R^2}-\cot\theta_0 \frac{x}{R}\, , ~  ~ ~ ~ ~ ~
~ ~ ~ \varphi= \frac{z}{R\sin\theta_0}\, . \label{cha} \eea
After this scaling and the limit $R\to \infty$, the metric becomes,
\bea
ds^2 &=& -4dudv + 4 y dudx - (r^2+z^2)du^2 \nonumber\\
&&+dx^2+dy^2+ dz^2 + z^2 d\xi^2 + dr^2 +r^2d\Omega_3^2 \, . \eea
If we define cartesian coordinates
\bea
&& z_1 = z\sin\xi\, , ~ ~ ~ ~ ~ ~ ~ ~ ~ ~ ~ ~ ~ ~ z_4= r \sin\varphi'\cos\eta'\, \nn\\
&& z_2 = z\cos\xi\, , ~ ~ ~ ~ ~ ~ ~ ~ ~ ~ ~ ~ ~ ~ z_5= r \cos\varphi'\sin\xi'\, , \nn\\
&& z_3= r \sin\varphi'\sin\eta'\, ,  ~ ~ ~ ~ ~ ~ z_6= r
\cos\varphi'\cos\xi'\, , \eea
the metric is written as,
\be ds^2 = -4dudv + 4 y dudx - \sum^6_{a=1} z_a^2 du^2 +dx^2+dy^2+
\sum^6_{a=1} dz_a^2  \, , \label{appw} \ee
and the RR 5-form field strength is
\be F_{(5)} = 2 du\wedge(dz_1\wedge dz_2 \wedge dz_3 \wedge dz_4 +
dz_5\wedge dz_6 \wedge dz_7 \wedge dz_8 )\, . \ee

This pp-wave configuration can be put in the standard form using the
following coordinate transformation \cite{Miche}
\bea
&& x_+= u\, , ~ ~ ~ ~ ~ ~ ~ ~ ~ ~ ~  ~ ~ ~ x_1 = x\cos u+y\sin u \, ,\nn\\
&& x_-= v- \frac{1}{2}xy\, , ~ ~ ~ ~ ~ x_2 = -x\sin u+y\cos u\, ,\nn\\
&& x_{a+2} = z_a\, , ~ ~ ~ ~ ~ ~ ~ ~ ~ ~  {\rm for} ~ ~
a=1,\ldots,6\,. \eea
which  leads the  metric to
\be ds^2 = -4dx_+dx_- - \sum_{i=1}^8 x_i^2 dx_+^2 + \sum_{i=1}^8
dx_i^2  \, . \ee
Notice that this transformation involves rotations at constant angular velocity 
in the $x,y$ plane. In this sense, the D-brane we will be considering are rotating in the standard plane wave limit.

\section{Brane in the Penrose limit}
\label{abpl} To see how the D3-brane is specified in terms of the
new coordinates (\ref{cha}) we should apply the coordinate
transformation to the boundary conditions and then take the
$R\to\infty$ limit. In the original coordinates, the giant expands
in $(\varphi, \eta,\xi)$, and then for an open string, they should
satisfy Neumann boundary conditions
\bea \label{bc1}
\left.\partial_\sigma\varphi\right|_{\sigma=0,\pi}&=&0\, , \\
\left.\partial_\sigma\eta\right|_{\sigma=0,\pi}&=&0\, ,   \\
\left.\partial_\sigma\xi\right|_{\sigma=0,\pi}&=&0\, . \eea
On the other hand, the giant is situated in $\rho=0$ and
$\theta=\theta_0$. Therefore, these coordinates have Dirichlet
boundary conditions
\bea \label{bc2}
\left.\delta\rho\right|_{\sigma=0,\pi}&=&0\, ,  \\
\left.\delta\theta\right|_{\sigma=0,\pi}&=&0\, . \eea
The remaining transverse direction to the D-brane is $\psi$ and
the D-brane is moving along this angle with $\psi=t$.  The
corresponding Dirichlet boundary condition is
\be \label{bc3} \left.\delta\psi\right|_{\sigma=0,\pi}  =
\left.\delta t\right|_{\sigma=0,\pi} \, . \ee
As a consequence, the remaining Neumann boundary condition also
mixes $t$ and $\psi$
\be \label{bc4} \left.\partial_\sigma t\right|_{\sigma=0,\pi}=
\cos^2\theta_0\left.\partial_\sigma\psi\right|_{\sigma=0,\pi}\, .
\ee
Using (\ref{cha}) the set of  boundary conditions
(\ref{bc1})-(\ref{bc4}) is translated, before the limit
$R\to\infty$, into
\bea \label{bbc1}
\left.\partial_\sigma z\right|_{\sigma=0,\pi}&=&0\, , \\
\left.\partial_\sigma\xi\right|_{\sigma=0,\pi}&=&0\, ,   \\
\left.\partial_\sigma v\right|_{\sigma=0,\pi}&=&0\, ,   \\
\left.\partial_\sigma u\right|_{\sigma=0,\pi}&=&
    \frac{\cot\theta_0}{R}\left.\partial_\sigma x\right|_{\sigma=0,\pi}\, ,   \\
\left.\delta\rho\right|_{\sigma=0,\pi}  &=& 0 \, ,\\
\left.\delta y\right|_{\sigma=0,\pi} &=& 0 \, ,\\
\left.\delta x\right|_{\sigma=0,\pi}  &=& \frac{2\cot\theta_0}{R}
      \left.\delta v\right|_{\sigma=0,\pi} \, .
\eea
So, in the $R\to\infty$ limit, we have  Neumann boundary conditions
for $u,v,z_1,z_2$ and Dirichlet boundary conditions for
$x,y,z_3,z_4,z_5,z_6$.


\section{Combinatorics and field theory calculations}
\label{con}

We begin by listing  some properties of the totally anti-symmetric
tensor that are useful for field theory  calculations. This tensor
is defined as
\begin{eqnarray}
\epsilon^{i_1\cdots i_p}_{j_1\cdots j_p}\equiv \left\{
\begin{array}{rll}
1 & & {\rm if\ } (i_1\cdots i_p)\ {\rm is\ an\ even\ permutation\ of\ } (j_1\cdots j_p)\\
-1 & & {\rm if\ } (i_1\cdots i_p)\ {\rm is\ an\ odd\ permutation\ of\ } (j_1\cdots j_p)\\
0 & &{\rm otherwise}
\end{array}\right.\; ,
\end{eqnarray}
where $p$ is any integer $p\leq N$, and $i_1,\cdots, i_p$ and $j_1,
\cdots, j_p$ are integers from $1$ to $N$.  The simplest examples
are
\begin{eqnarray}
\epsilon^{i}_j&=&\delta^{i}_{j}\;,\nonumber\\
\epsilon^{ij}_{kl}&=&\delta^{i}_{k}\delta^{j}_{l}-\delta^{i}_{l}\delta^{j}_{k}
\;.
\end{eqnarray}

Some  of most useful properties of the $\epsilon$ tensor are
\begin{eqnarray}
 \label{formula1}
\epsilon^{i_1\cdots i_p}_{j_1\cdots
j_p}&=&\sum_{x=1}^{p}(-1)^{x+1}\delta^{i_1}_{j_x}
\epsilon^{i_2\;\;\;\;\;\; .\;\, .\;\, .\;\;\;\;\;\;\;
i_p}_{j_1\cdots j_{x-1}j_{x+1}\cdots j_p}\;, \\
 \label{formula2}
\epsilon^{i_1\cdots i_k i_{k+1}\cdots i_p}_{i_1\cdots i_k
j_{k+1}\cdots j_p}&=&\frac{(N-p+k)!}
{(N-p)!}\epsilon^{i_{k+1}\cdots i_p}_{j_{k+1}\cdots j_p} \;,\\
 \label{formula3}
\epsilon^{i_1\cdots i_k}_{j_1\cdots j_k}\epsilon^{j_1\cdots
j_p}_{l_1\cdots l_p}&=&k!\epsilon^{i_1\cdots i_k j_{k+1}\cdots
j_p}_{l_1\;\;\;\;\; .\;\, .\;\, .\;\;\;\;\;\; l_p}\;.
\end{eqnarray}
Another useful identity that can be derived from the ones above is
\begin{eqnarray}
\label{identity}
 \epsilon_{i_1\cdots i_{p-1} \mu_1}^{j_1 \cdots
j_{p-1} \gamma_1} \epsilon_{j_1 \cdots j_{p-1} \mu_2 \cdots
\mu_k}^{i_1 \cdots i_{p-1} \gamma_2 \cdots \gamma_k} = \frac{(p-1)!
(N - k + 1)!}{(N - p - k + 2)!} \left(\delta_{\mu_1}^{\gamma_1}
\epsilon_{\mu_2 \cdots \mu_k}^{\gamma_2 \cdots \gamma_k}  -
\frac{(p-1)}{(N-k + 1)} \epsilon_{\mu_1 \cdots \mu_k}^{\gamma_1
\cdots \gamma_k}\right) \;. \nonumber \\
\end{eqnarray}

In the rest of this appendix  we present some of the field theory
calculations
 leading to the Hamiltonian (\ref{H}). Most of the
calculations are very similar to those presented in \cite{sam} in
the context of maximal giant gravitons.

We begin by pointing out that an operator with a $Z$ field at the
end or beginning of the word does not represent a linearly
independent state. In fact, from the identities above one can show
that
\begin{eqnarray} \epsilon_{i_1 \cdots i_p}^{j_1 \cdots j_p}
Z_{j_1}^{i_1} \cdots Z_{j_{p-1}}^{i_{p-1}} (Z W)_{j_p}^{i_p} =
 \frac{1}{p} \epsilon_{i_1 \cdots i_p}^{j_1 \cdots j_p} Z_{j_1}^{i_1} \cdots
Z_{j_p}^{i_p} \tr(W) - \frac{1}{p} \epsilon_{i_1 \cdots
i_{p+1}}^{j_1 \cdots j_{p+1}} Z_{j_1}^{i_1} \cdots
Z_{j_{p}}^{i_{p}}W_{j_{p+1}}^{i_{p+1}}\;. \nonumber \\
\end{eqnarray}
This shows that the operator with a $Z$ at the border of the word
 factorizes into a brane with a closed
string and a bigger brane.

Now we turn our attention to the correlation functions. These will
be computed to one loop in the 't Hooft coupling and using the
planar approximation of the large $N$ limit. We will be using the
bosonic part of the SYM action given by
\begin{eqnarray}
\label{action} S = \frac{1}{2 \pi g_s} \int d^4x
\textrm{Tr}\left(\frac{1}{2} F_{\mu \nu} F^{\mu \nu} + D_\mu X D^\mu
\overline{X} + D_\mu Y D^\mu \overline{Y} + D_\mu Z D^\mu
\overline{Z} + V_D + V_F\right), \nonumber \\
\end{eqnarray}
where,
\begin{eqnarray}
V_D &=& \frac{1}{2} \textrm{Tr}\left( | [X, \overline{X}] + [Y,
\overline{Y}] + [Z, \overline{Z}] |^2 \right)\;, \\
V_F &=& 2 \textrm{Tr}\left( |[X,Y]|^2 + |[X,Z]|^2 + |[Y,Z]|^2
\right)\;.
\end{eqnarray}

Let us begin with the free field correlation functions, since they
are going to give us the normalization of the operators.  The
propagator for the scalars is of the form
\begin{equation}
\bra \phi^i_j(x) \bar \phi^k_l(0) \ket = \frac{g_s}{2 \pi}
\frac{1}{| x |^2}\delta^i_l \delta^k_j\;,
\end{equation}
where $\phi$ is any of the complex scalars $X, Y, Z$. Thus we see
that the correlation function of any word with a total classical
dimension $\Delta_0$ will be multiplied by the following overall
numerical factor: $ (\frac{g_s}{2\pi}\frac{1}{|x|^{2
}})^{\Delta_0}$. In what follows we will drop this factor from the
calculations for simplicity.

Let us now compute the correlation function for the operator
\begin{equation}
{\cal O}^p_W  = \epsilon_{i_1 \cdots i_p}^{j_1 \cdots j_p}
Z_{j_1}^{i_1} \cdots Z_{j_{p-1}}^{i_{p-1}} W_{j_p}^{i_p}\;,
\end{equation}
where the word $W$ has classical dimension $L$. We will assume that
the operator is very long such that $p > \sqrt{N}$ and $L \lesssim
\sqrt{N}$ with $p \gg L$. In the large $N$ limit, the leading
contribution will come from contracting all the $Z$ fields and then
the words planarly. Moreover, in the large $N$ limit operators with
different values of $p$ and word lengths will be orthogonal. Thus the free
theory two point function gives
\begin{eqnarray}
\label{norm} \bra \bar{{\cal O}}^p_W {\cal O}^p_{W}\ket_{\text free}
&\sim& (p - 1)! \epsilon_{i_1 \cdots i_{p-1} \mu_1}^{j_1 \cdots
j_{p-1} \gamma_1} \epsilon_{j_1 \cdots j_{p-1} \mu_2}^{i_1 \cdots
i_{p-1} \gamma_2} \bra\! \bar{W}_{\gamma_1}^{\mu_1}
W_{\gamma_2}^{\mu_2}
\ket_{\text free} \nonumber \\
 &=& \frac{(p-1)!^2 (N- 2)!}{(N-p-1)!}\left[
 \bra\!\tr(\bar W) \tr(W)\ket_{\text free} + \frac{(p-1)}{N-p} \bra\!\tr({\bar W} W)\ket_{\text free}
 \right] \nonumber \\
 &\sim& \frac{p! (p - 1)! (N - 1)! N^L }{(N - p)!}\;.
\end{eqnarray}
In going to the  second line of (\ref{norm}) we used the identity
(\ref{identity}) and for the third line we contracted the words
using the planar approximation.

If the word $W$ is made of a single type of field we will get an
additional multiplicative factor of $L(1 + {\cal O}(L/N))$ to
(\ref{norm}) from the cyclic property of the trace.  This is one of
the reasons we consider only $L \ll N$ and in particular $L \lesssim
\sqrt{N}$.

We now turn on the Yang-Mills coupling and consider the interactions
between the scalars. Let us start by calculating the interactions in
the bulk of the words. This terms are obtained when all the $Z$
fields of the giants are contracted we free field propagators and
the vertex is entirely contracted with letters of the words. In the
planar approximation these will be the familiar nearest-neighbor
interactions that were considered by Minahan and Zarembo
\cite{minahan}. Thus if we regard each ``letter" in $W$ as a $SO(6)$
vector, the nearest-neighbor interactions have the familiar form,
\begin{equation}
\label{bulkH}
 H_{l,l+1} = \frac12 \lambda [ K_{l,l+1} +
2 (I_{l,l+1} - P_{l,l+1})] \;,
 \end{equation}
 where $K$ and $P$ are the trace and permutation operators
 respectively and $\lambda = g_s N/ {2 \pi}$. For the $SU(2)$ and $SU(3)$
sectors considered above, the trace will be zero and it is easy to
see that the remaining interactions account for the first two terms
in the Hamiltonian (\ref{H}).

 For completeness, let
 us show how to calculate the bulk part of the Hamiltonian by
 considering the example of the interactions giving rise to the
 identity in (\ref{bulkH}). To this end we  study the interaction
 between a $X$ and $Y$ fields of the operator
 \begin{eqnarray}
 \label{opxy}
 {\cal O}^p_{XY} =\epsilon_{i_1 \cdots i_p}^{j_1 \cdots j_p}
Z_{j_1}^{i_1} \cdots Z_{j_{p-1}}^{i_{p-1}} (W_1 X Y
W_2)_{j_p}^{i_p}\;.
\end{eqnarray}
It is well known that for chiral operators the interactions coming
from $V_D$ cancel with the gauge boson exchange and the scalar self
energies. Therefore, the only relevant interactions between $X$ and
$Y$ will come from the F term. It is not difficult to see that the
only {\it planar} interaction will be,
\begin{equation}
-\frac{ 1}{\pi g_s} \int d^4x \tr(XY\bar{Y}\bar{X})\;.
\end{equation}
This interaction is illustrated in figure \ref{fig:bulkdiag.eps}.

\myfig{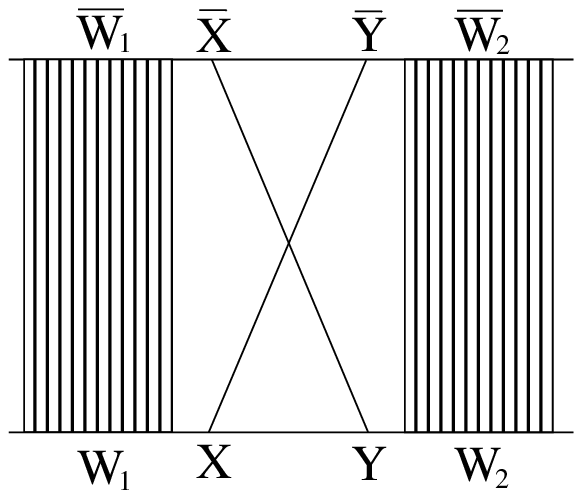}{4}{One-loop interaction between two nearest-neighbor
letters of a word generated by a vertex of the form $\tr(XY\bar Y\bar X)$.}

The rest of the free theory contractions go as before. Therefore,
all we get from the interaction is a multiplicative correction to
(\ref{norm}) of the form
\begin{eqnarray}
\left(\frac{g_s}{2 \pi}\right)^{-2}\left( \frac{g_s}{2\pi} \right)^4
N \left(- \frac{1}{\pi g_s}\right)  |x|^4 \int d^4y \frac{1}{|y|^4|x
- y|^4} \approx  - 2 \lambda \log(|x| \Lambda)\;.
\end{eqnarray}
Thus, after dividing by the norm (\ref{norm}) and using the
definition of the anomalous dimension matrix (\ref{Mexp}), we see
that this interaction will give a numerical contribution of
$\lambda$ to the identity interaction in agreement with
(\ref{bulkH}).

Now let us see what happens when we have a non-chiral operator. Let
us consider the word $W_1 X \bar Y W_2$.  In this case the D-term
interaction will no longer cancel with the gauge boson exchange and
the scalar self-energies. However, we can exploit the fact that
these last two interactions are flavor blind to write their
contribution in terms of the D-term contribution to the chiral
operator. This is illustrated in figure \ref{fig:diagram1.eps}.

\myfig{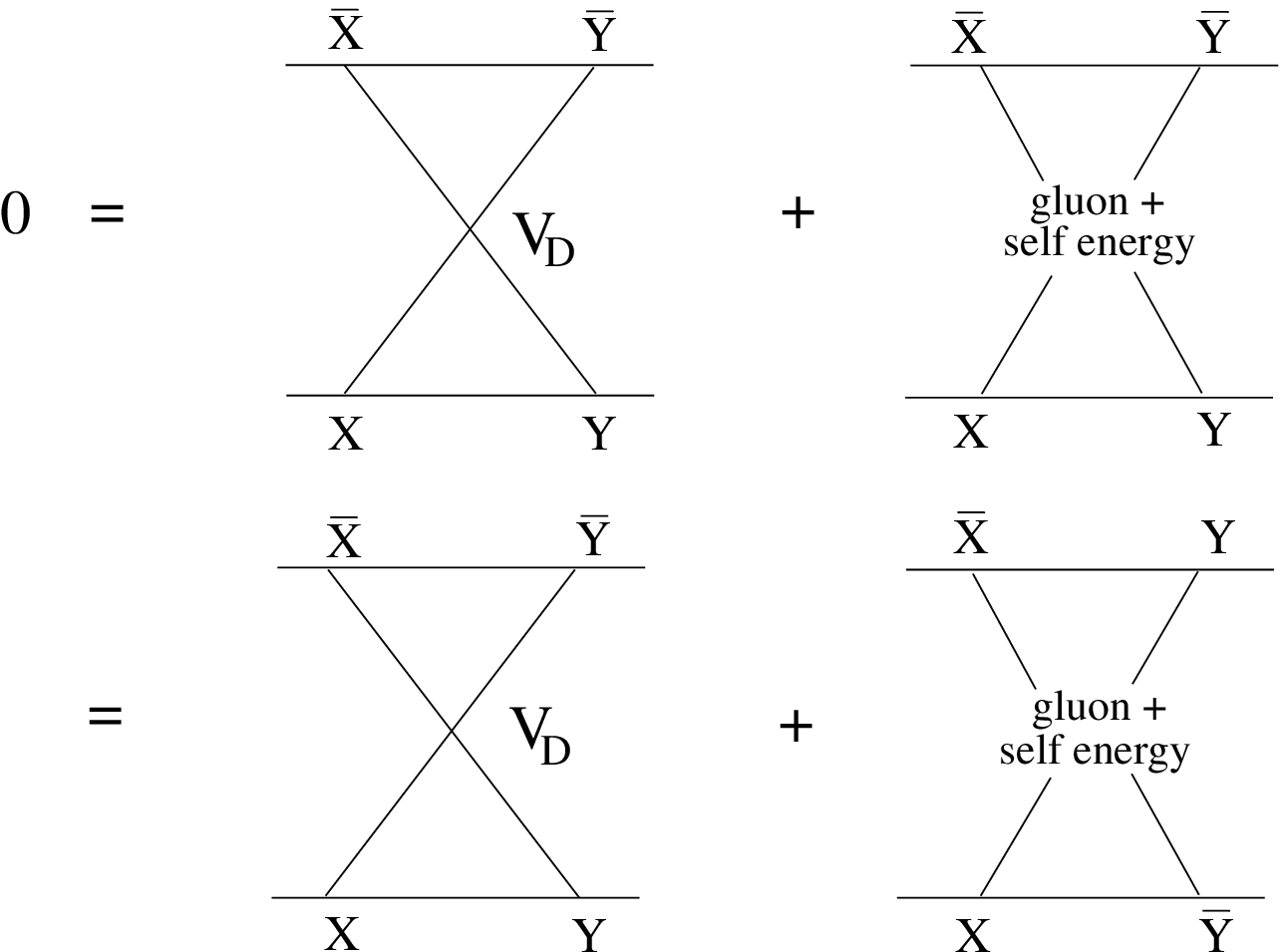}{8}{The first line of the figure illustrates the
cancelation of the D-term contribution with the gluon exchange and
the scalars self energies for a chiral operator. Since the gluon and
self energy interactions are flavor blind they will give the same
numerical value if we compute them using the non-chiral operator as
in the second line.}

 It is not difficult to see that there
is no planar contribution to the identity coming from the F-term.
Thus, the total contribution to this interaction is from the
diagrams shown in figure \ref{fig:diagram2.eps}.

\myfig{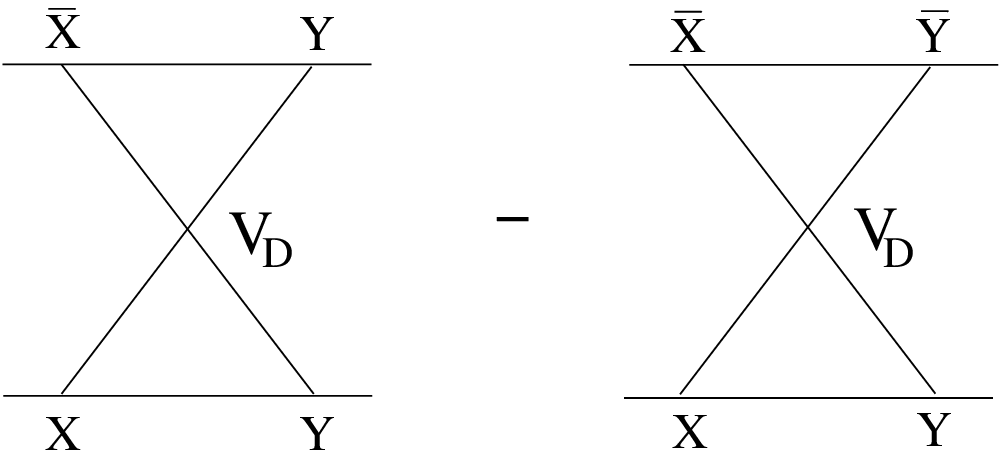}{6}{{Total contribution to the one-loop interaction between $X$ and $\bar Y$ of the word
$W_1 X \bar Y W_2$. The second term is the gluon and self energy interactions calculated
in terms of $V_D$ using the result illustrated in the previous figure. }}

The first diagram comes from the vertex
\begin{equation}
-\frac{1}{2 \pi g_s} \tr(X\bar Y Y \bar X)\;,
\end{equation}
and the second from
\begin{equation}
-\frac{1}{2 \pi g_s} \tr(X Y \bar Y \bar X)\;.
\end{equation}
Note that the second vertex is to be evaluated between the chiral
operator. At the end, we see that we will get a numerical
contribution of $\lambda /2$ from each interaction, and this will
sum up to the contribution shown in (\ref{bulkH}). The rest of the
interactions in the Hamiltonian (\ref{bulkH}) can be obtained in a
similar way.

Let us now turn our attention to the interactions between the word
and the $Z$ fields of the giant graviton. The first contribution
comes from the scalars at the ends of the word. Thus we can study
the operator
\begin{eqnarray}
 {\cal O}^p_{\phi} =\epsilon_{i_1 \cdots i_p}^{j_1 \cdots j_p}
Z_{j_1}^{i_1} \cdots Z_{j_{p-1}}^{i_{p-1}} (\phi W)_{j_p}^{i_p}\;,
\end{eqnarray}
where $\phi$ is any of the complex scalars except $Z$ and the word
$W$ has length $L -1$. Doing the free contractions we get
\begin{eqnarray}
\label{corrborder} \bra \bar{\cal O}^p_\phi (x) {\cal O}^p_\phi(0)
\ket &\sim& N^{L-2} (p-1)^2 (p-2)! \epsilon_{i_1 \cdots i_{p-2}
\mu_1 \mu_2}^{j_1 \cdots j_{p-2} \gamma_1 \gamma_2}\epsilon_{j_1
\cdots j_{p-2} \mu_3 \gamma_2}^{i_1 \cdots i_{p-2} \gamma_3
\gamma_4} \nonumber \\
&& \times \bra \bar Z^{\mu_1}_{\gamma_1}(x) Z_{\gamma_3}^{\mu_3}(0)
[\phi(0) \bar
\phi(x)]_{\gamma_4}^{\mu_2} \ket \;. \nonumber \\
\end{eqnarray}

We now note that there are only two relevant planar interactions.
These will have the general form:
\begin{equation}
\label{ints} -\frac{\alpha}{\pi g_s} \tr(Z\bar Z \phi \bar
\phi)\;,\;\;\;\;\; -\frac{\beta}{\pi g_s} \tr(\bar Z Z \phi \bar
\phi)\;.
\end{equation}
If the field $\phi$ is $\bar X, \bar Y$ or $\bar Z$, the constants
$\alpha$, $\beta$ will need to include the contribution from the
gauge boson exchange and the scalar self energies. In any case, the
interactions (\ref{ints}) will give a contribution to
(\ref{corrborder}) of
\begin{eqnarray}
\bra \bar{\cal O}^p_\phi (x) {\cal O}^p_\phi(0) \ket  \sim -  2
\lambda N^{L} \frac{p! (p-1)! (N-1)!}{(N-p)!} \left[ \alpha
\frac{p}{N} + \beta
\left(1 - \frac{p}{N}\right) \right] \log (|x| \Lambda) \;,\nonumber \\
\end{eqnarray}
up to multiplicative corrections of order $1/p$ and $1/N$. Thus
after dividing by the norm (\ref{norm})  we get the following
contribution to the anomalous dimension: $ \lambda \left[ \alpha
p/N + \beta \left(1 - p/N\right) \right] $. Calculating the
constants $\alpha$ and $\beta$ is just a matter of counting how many
terms like (\ref{ints}) we can find in $V_D$ and $V_F$. At the end
we get that for the fields $X, Y, \bar X$ or $\bar Y$, $\alpha = 0$
and $\beta = 1$. Thus the anomalous dimension gives $\lambda (1 -
p/N)$. Translating to the boson lattice language and taking into
account the contribution to the field at the other end of the word,
this gives the first term in the second line of the Hamiltonian
(\ref{H}). On the other hand, if $\phi = \bar Z$ we have that
$\alpha = -1$ and $\beta = 3/2$. Thus the contribution to the
anomalous dimension is $\lambda (1 + (1 - p/N)/2)$.

Finally, there is an interaction that is not present in the case of
the maximal giant graviton. This is the exchange of a $Z$ field
between  the word and the giant graviton. In the planar
approximation this can only happen when the $Z$ is  the second or
next to last letter of the word. To calculate this amplitude we can
consider the correlation between the operators
\begin{eqnarray}
 {\cal O}^{p+1} =\epsilon_{i_1 \cdots i_{p+1}}^{j_1 \cdots j_{p+1}}
Z_{j_1}^{i_1} \cdots Z_{j_{p}}^{i_{p}} (X W)_{j_{p+1}}^{i_{p+1}}\;,
\;\; {\cal O}^{p} =\epsilon_{i_1 \cdots i_{p}}^{j_1 \cdots j_{p}}
Z_{j_1}^{i_1} \cdots Z_{j_{p-1}}^{i_{p-1}} (X Z
W)_{j_{p}}^{i_{p}}\;,
\end{eqnarray}
and assume that the word $W$ has classical dimension $L$. The
calculation here is very similar to the previous example. The free
contractions give
\begin{eqnarray}
 \bra \bar{\cal O}^{p+1} (x) {\cal O}^p(0)
\ket &\sim& N^{L-1} p! \epsilon_{i_1 \cdots i_{p-1} \mu_1
\mu_2}^{j_1 \cdots j_{p-1} \gamma_1 \gamma_2}\epsilon_{j_1 \cdots
j_{p-1}  \gamma_2}^{i_1 \cdots i_{p-1} \gamma_3}
\bra \bar Z^{\mu_1}_{\gamma_1} (X Z \bar X)^{\mu_2}_{\gamma_3}  \ket \;. \nonumber \\
\end{eqnarray}
Since these are chiral operators, the only interactions will come
from $V_F$. It is easy to see that the only planar interaction  will
be
\begin{equation}
+\frac{1}{\pi g_s} \tr(Z X \bar Z \bar X)\;.
\end{equation}
Doing the contractions we get
\begin{eqnarray}
 \bra \bar{\cal O}^{p+1} (x) {\cal O}^p(0)
\ket &\sim&  2\lambda N^{L} p! \epsilon_{i_1 \cdots i_{p-1} \mu_1
\gamma_1}^{j_1 \cdots j_{p-1} \gamma_1 \gamma_2}\epsilon_{j_1
\cdots j_{p-1}  \gamma_2}^{i_1 \cdots i_{p-1} \mu_1} \log(|x| \Lambda) \nonumber \\
&=& -2 \lambda N^{L+2} \frac{p!^2 (N-1)!}{(N-p)!} \left(1 -
\frac{p}{N}\right) \log(|x| \Lambda)\;,
\end{eqnarray}
where we have used the identities (\ref{formula2}) and
(\ref{formula3}).

The norms are
\begin{eqnarray}
\bra \bar {\cal O}^p {\cal O}^p\ket_{\text free} &\sim& \frac{p!^2
(N-1)! N^{L+2}}{p (N-p)!}\;, \\
\bra \bar {\cal O}^{p+1} {\cal O}^{ p+1}\ket_{\text free} &\sim&
\frac{(p+1) p!^2 (N-1)! N^{L+2}}{ (N-p)!}\left(1 -
\frac{p}{N}\right)\;.
\end{eqnarray}
Therefore, for large $p$ the contribution to the anomalous dimension
is: $\lambda \sqrt{1 - p/N}$. We recognize this as sources/sinks in
the Hamiltonian (\ref{H}).

To conclude we show that closed string emission/absorption is
suppressed in the large $N$ limit if we take $p \sim \gamma N$. For
definitiveness consider the operators
\begin{eqnarray}
{\cal O}^p_1 =  \epsilon_{i_1 \cdots i_p}^{j_1 \cdots j_p}
Z_{j_1}^{i_1} \cdots Z_{j_p}^{i_p} \tr(Y^L)\;,\;\;\;\; {\cal O}^p_2
= \epsilon_{i_1 \cdots i_p}^{j_1 \cdots j_p} Z_{j_1}^{i_1} \cdots
Z_{j_{p-1}}^{i_{p-1}} (YZY^{L-1})_{j_p}^{i_p}\;.
\end{eqnarray}
Then, the interacting correlation function gives (up to signs or
numerical factors)
\begin{eqnarray}
\bra \bar {\cal O}^p_1 {\cal O}^p_2\ket \sim \lambda \frac{L p!^2
(N-1)! N^L}{(N-p)!}\log(|x| \Lambda)\;,
\end{eqnarray}
The norms are,
\begin{eqnarray}
\bra \bar{\cal O}_1 {\cal O}_1\ket_{\text{free}} \sim  \frac{L p!^2
N! N^L}{(N-p)!}\;,\;\;\; \bra \bar {\cal O}_2 {\cal
O}_2\ket_{\text{free}} \sim \frac{ p!^2 N!
N^L}{p (N-p)!}\;,\nonumber\\
\end{eqnarray}
Therefore, the contribution to the anomalous dimension is of the
order
\begin{eqnarray}
 \lambda \frac{ \sqrt{L p}}{N} \sim \lambda \sqrt{\frac{L}{N}}\;,
 \end{eqnarray}
This is clearly suppressed in the limit we are considering of $L
\lesssim \sqrt{N}$.

\section{Calculations in the $SU(3)$ sector}
\label{su3app}

Here we give some more details about the calculation of the effective sigma model action for the $SU(3)$
sector. We begin with the gauge theory side. The coherent states for states of the form (\ref{su3word})
take the product form
\begin{equation}
|\text{CS}\ket = \bigotimes_{i = 1}^L | \vec{n}_i, z_i\ket \otimes |
\vec{n}_{L+1}\ket\;,
\end{equation}
where $|z_i\ket$ are the bosonic coherent states (\ref{coherent}) and $|\vec{n}_i\ket$ are the $SU(2)$
coherent states defined by  \cite{zhang,Kruczenski},
\begin{equation}
|\vec{n} \ket =  e^{i S_z \varphi} e^{i S_y \theta} |\uparrow
\ket\;,
\end{equation}
where $\vec{n} = (\sin \theta \cos \varphi, \sin \theta \sin
\varphi, \cos\theta)$ is a unit vector in $S^2$. They obey \be \bra \vec
n | \vec{S} | \vec{n} \ket = \frac{1}{2} \vec{n}\;, \ee \be \bra
\vec n |\vec n' \ket = \left[ \cos \frac{1}{2} (\theta -
\theta')\cos \frac{1}{2} (\varphi - \varphi') - i \cos\frac{1}{2}
(\theta + \theta') \sin\frac{1}{2} (\theta - \theta')\right] e^{i
\frac{1}{2}(\varphi - \varphi')}\;. \ee

The action for the coherent states is
\begin{eqnarray}
S &=& \lim_{L \rightarrow \infty} \int dt \left(i \bra \text{CS}|
\partial_t | \text{CS}\ket -\bra \text{CS}|
H | \text{CS}\ket\right) \;,
\end{eqnarray}
and we get (\ref{Sthree}).

On the other hand, one can study the dual string theory and write
the Polyakov action for open strings as we did in section 2. In the
case of three spins, the natural coordinates on the sphere $|X|^2 +
|Y|^2 + |Z|^2 = 1$ are
\begin{eqnarray}
\label{su3coor} X = \pm \sqrt{1  - r^2} \cos\theta\, e^{i
\varphi_1}\;,\;\;\; Y = \pm \sqrt{1 - r^2}  \sin\theta\, e^{i
\varphi_2}\;,\;\;\; Z = r e^{i(t - \phi)}\;,
\end{eqnarray}
where the giant graviton is located at $r = \sqrt{1 - p/N}$ and
$\varphi = \varphi_1 + \varphi_2 - t$.  Following \cite{kru} we can
make another change of coordinates:
\begin{equation} \alpha = \frac{1}{2}
(\varphi_1 + \varphi_2)\;, \;\;\; \beta = \frac{1}{2} (\varphi_1 -
\varphi_2)\;, \;\;\; U_1 = \cos \theta e^{i \beta}\;, \;\;\; U_2 =
\sin \theta e^{-i \beta}\;.
\end{equation}
In these coordinates, the metric on $\mathbb{R} \times S^5$ takes
the form
\begin{equation}
ds^2 = - (1 - r^2) dt^2 + \frac{1}{1 - r^2} dr^2 + r^2 d\phi^2 + 2
r^2 d\phi dt + (1 - r^2)\left[ (D\alpha)^2 + \frac{1}{4} dn_i
dn_i\right]\;,
\end{equation}
where coordinates on a $S^2$ are given by $n_i =
\xi^\dagger \sigma_i \xi$, where $\xi = (U_1 e^{i \alpha}, U_2 e^{-i
\alpha})$ and $\sigma_i$ are the Pauli matrices. The covariant
derivative is defined as $D \alpha = d\alpha + C$, where $C = -i
\bar U_i dU_i$.

We now want to expand the Polyakov action in the limit of large
angular momentum in $\alpha$, and choose a gauge where this angular
momentum is uniformly distributed along the string. Note that this
angular momentum is dual to the total number of $X$s and $Y$s in the
word (\ref{su3word}). We choose the coordinates in the Polyakov
action such that $\alpha$ is coupled through its covariant
derivatives $\sim g^{a b} D_a \alpha D_b \alpha$, where $D_a \alpha
= \partial_a \alpha
 - i\bar U_i \partial_a U_i$.

 With this choice, the lagrangian density
 in momentum space (\ref{paction}) becomes
\begin{equation}
\label{lagrangian} {\cal L} = p_t \dot t + p_r \dot r + p_\phi \dot
\phi + p_\alpha D_t \alpha + p_i \dot n_i\;,
\end{equation}
and we have the usual constraints:
\begin{eqnarray}
\label{const1}
G^{\mu \nu} p_\mu p_\nu + G_{\mu \nu} X'^\mu X'^\nu &=& 0\;,\\
\label{const2} p_\mu X'^\mu &=& 0 \;.
\end{eqnarray}

We then choose the usual gauge: $p_\alpha = 2 {\cal J} =
\text{constant}$ and $t = \tau$, where the total angular momentum in
$\alpha$ is $L = \sqrt{\lambda_{YM}} {\cal J}$. From (\ref{const2})
we can eliminate $D_1 \alpha$ and then use (\ref{const1}) to
eliminate $p_t$ from the Lagrangian (\ref{lagrangian}). The result
is
\begin{equation}
{\cal L} = p_\phi + p_a \dot X^a + 2 {\cal J} C_0 - \sqrt{\Lambda^{a
b} p_a p_b + \Lambda }\;,
\end{equation}
where the indices $a, b$ run trough the rest of the coordinates
excluding $t$ and $\alpha$. The functions $\Lambda^{a b}$ and
$\Lambda$ are
\begin{eqnarray}
\Lambda &=& \frac{4 {\cal J}^2}{1 - r^2} + \frac{r'^2}{1 - r^2} +
r^2
\phi'^2 + (1 - r^2) n_i'^2\;, \\
\Lambda^{rr} &=& (1 - r^2) \left(1 + \frac{r'^2}{4 {\cal J}^2}
\right)\;,\\
\Lambda^{\phi \phi} &=& 1 + (1 - r^2) \left(\frac{1}{r^2} +
\frac{\phi'^2}{4 {\cal J}^2} \right)\;, \\
\Lambda^{i i } &=& \frac{1}{1 - r^2} + \frac{1 - r^2}{4 {\cal J}^2}
{n'}_i^2 \;, \\
\Lambda^{r \phi} &=&  \Lambda^{ \phi r}  = \frac{(1 - r^2) r' \phi'}{4 {\cal J}^2}\;, \\
\Lambda^{r i} &=& \Lambda^{i r} =  \frac{(1 - r^2) r' n_i'}{4 {\cal J}^2}\;, \\
\Lambda^{\phi i} &=&  \Lambda^{ i \phi} = \frac{(1 - r^2) \phi' n_i'}{4 {\cal J}^2}\;, \\
\Lambda^{i j} &=&  \Lambda^{j i} = \frac{(1 - r^2) n_i' n_j'}{4 {\cal J}^2}\;. \\
\end{eqnarray}
We then proceed as before by eliminating the momenta using their
equations of motion. This gives
\begin{equation}
{\cal L} = 2 {\cal J} C_0 - \sqrt{\Lambda(1 - \Lambda_{a b} \dot x^a
\dot x^b)}\;,
\end{equation}
where $\dot x^a = \dot X^a$ for $a \neq \phi$ and $\dot x^\phi = 1 +
\dot \phi$, and $\Lambda_{a b}$ is the inverse of the symmetric
matrix $\Lambda^{a b}$.

One can now expand as usual in the limit of large $\cal J$ assuming
$\dot X^a \sim {\cal O}(1/{\cal J}^2)$. To lowest non-trivial order
we get the action (\ref{Sthree}) except that for the $r$ dependence
of the last term we get $\int d^2\sigma (1 - r^2)^2 (\partial_1
n_i)^2$.

\section{Coherent states for the $q$-deformed algebra}
\label{qch}

The $q$-deformed algebra has the following representation
\cite{arik}, \bea \label{defops} a^\dagger |n\rangle \!=\!
\sqrt{[n+1]}\ |n+1\rangle \, ,\;\;\; a |n\rangle \!=\! \sqrt{[n]}\
|n-1\rangle\, , \;\;\;{\rm with}\;\; [n] \equiv \frac{\ 1-q^n}{1-q}
\,. \eea Coherent states can  be defined as eigenstates of the
annihilation operator, and they are given by \cite{baz}
\be  \label{coherent} |z\rangle =\left( \exp_q(|z|^2)\right)^{-1/2}
\sum_{n=0}^{\infty} \frac{z^n}{\sqrt{[n]!}}|n\rangle\,, \;\;\;\;\;\;
{\rm with}\;\;\;|z|<\frac{1}{1-q}\; , \ee
where $[n]!= [1].[2] \ldots [n]$. The $q$-exponential is defined by,
\be \exp_q(x) =\sum_{n=0}^{\infty} \frac{x^n}{[n]!} =
\frac{1}{\Pi_{k=0}^\infty (1 - q^k(1 -q) x)}\,. \ee
As usual, these coherent states constitute a non-orthogonal and
overcomplete basis. The overlapping between coherent states is
\be \bra z| z'\ket = \frac{\exp_q(\bar z
z')}{\left(\exp_q(|z|^2)\exp_q(|z'|^2)\right)^{1/2}}\;. \ee
The resolution of the identity takes the form,
\begin{eqnarray}
\frac{1}{\pi} \int_{D_q} d^2_q z |z\ket\bra z| = \frac{1}{2\pi}
\int_0^{2\pi} d\phi \int_0^{1/(1-q)^2} d_q(r^2) |z \ket\bra z| =
1\;.
\end{eqnarray}
where $D_q$ is the open disk in the complex plane with radius
$1/(1-q)$, $z = r e^{i\phi}$  and the integral over $r$ is the
so-called Jackson Integral,
\begin{eqnarray}
\int_0^a f(x) d_q x = a(1-q)\sum_{n = 0}^\infty q^k f(q^k a)\;,
\end{eqnarray}
which satisfies $\int_0^{x_1}(\exp_q(x))^{-1} x^n d_q x = [n]!$.

\end{document}